\documentclass[journal,twocolumn,10pt,twoside]{IEEEtranTCOM}
\usepackage[pdftex]{graphicx}
\usepackage{caption}
\usepackage{amsmath}
\usepackage{amssymb}
\usepackage{extarrows}
\usepackage{dsfont}
\usepackage{algorithm}
\usepackage{algpseudocode}
\usepackage{exscale,relsize}
\usepackage{flushend}
\usepackage{filecontents}
\usepackage{nicefrac}
\usepackage{hyperref}
\usepackage{cite}
\usepackage{subfigure}
\usepackage[table]{xcolor}
\usepackage{gensymb}
\usepackage{array}
\usepackage{longtable}
\usepackage{float}
\usepackage{textcomp}

\begin{document}

%
\title{WiMesh: Leveraging Mesh Networking For Disaster Communication in Poor Regions of the World}

\author{Usman Ashraf$^{1}$, Amir Khwaja$^{2}$, Junaid Qadir$^{3}$, Stefano Avallone$^{4}$, Chau Yuen$^{5}$
\thanks{$^{1}$U. Ashraf is with the college of Computer Science and Information Technology, King Faisal University, Saudi Arabia
        {\tt\small uashraf@kfu.edu.sa}}%
\thanks{$^{2}$A. Khwaja is with the Department of Information Systems, King Faisal University, Saudi Arabia
        {\tt\small akhwaja@kfu.edu.sa}}
\thanks{$^{3}$J. Qadir is with the Department of Electrical Engineering at the Information Technology University (ITU) in Lahore, Pakistan  {\tt\small junaid.qadir@itu.edu.pk}}%
\thanks{$^{4}$S. Avallone is with the Department of Computer Engineering at the University of Napoli Federico II, Italy  {\tt\small stavallo@unina.it}}
\thanks{$^{5}$C. Yuen is with Department of Engineering Product Development (EPD), Singapore University of Technology and Design, Singapore 138682 {\tt\small yuenchau@sutd.edu.sg}}%
}

\maketitle

\begin{abstract}
This paper discusses the design, implementation and field trials of WiMesh - a resilient Wireless Mesh Network (WMN) based disaster communication system purpose-built for underdeveloped and rural parts of the world. Mesh networking is a mature area, and the focus of this paper is not on proposing novel models, protocols or other mesh solutions. Instead, the paper focuses on the identification of important design considerations and justifications for several design trade offs in the context of mesh networking for disaster communication in developing countries with very limited resources. These trade-offs are discussed in the context of key desirable traits including security, low cost, low power, size, availability, customization, portability, ease of installation and deployment, and coverage area among others. We discuss at length the design, implementation, and field trial results of the WiMesh system which enables users spread over large geographical regions, to communicate with each other despite the lack of cellular coverage, power, and other communication infrastructure by leveraging multi-hop mesh networking and Wi-Fi equipped handheld devices. Lessons learned along with real-world results are shared for WiMesh deployment in a remote rural mountainous village of Pakistan, and the source code is shared with the research community.
\end{abstract}


\section{Introduction}

\IEEEPARstart{W}{}hile natural disasters have long been a scourge for humanity, our ability to respond to such emergencies has gradually been improving with the aid of technology. In recent times, a new powerful trend of ``digital humanitarism'' \cite{meier2015digital}  is emerging in which digital communication facilities are being used along with big data and crowdsourcing techniques to scale up humanitarian response to natural disasters and other crisis situations \cite{qadir2016crisis}. The availability of communication services in crisis-hit situations is essential due to the facility it provides in coordination between the various stakeholders. 

Wireless networking in particular has revolutionized communication by offering innovative, flexible, and cost-effective untethered solutions. Due to the significant benefits of the wireless technology, today we have a diverse range of wireless technologies at our disposal. Cellular technologies can be perhaps at the forefront of this wireless revolution and have penetrated even remote areas around the world. However, there are still some scenarios where these technologies fail. In particular, the cellular infrastructure as well as the existing Public Switched Telephone Network (PSTN) may be destroyed during disasters and alternate means may need to be investigated for disaster communication. Moreover, despite its deep penetration, there are still sizable areas of population at remote locations where the low return on investment (ROI) along with the difficulty of terrain have made deployment of cellular infrastructure infeasible. In addition, cellular technologies have also found it difficult to break into rural and low-income regions because common models of operations require a high Average Revenue Per User (ARPU), which rural areas cannot support since rural areas are sparsely populated \cite{onireti2016will}.

South-East Asia, and Pakistan in particular, is a disaster-prone area that has suffered a number of calamities that have wreaked havoc and created misery. This research work is particularly motivated by the increasing frequency of disasters in Pakistan in the last decade or so. On October 8th, 2005, there was the Kashmir earthquake that resulted in 80,000 fatalities and 3.5 million homeless people \cite{KashmirEarthquake}\cite{KashmirEarthquake2}. On October 29th, 2008, the Ziarat earthquake hit the Pakistan province of Balochistan resulting in 215 fatalities and 120,000 homeless people \cite{ZiaratEarthquake}\cite{PakistanEarthquake}. During the earthquakes, the traditional telecommunication infrastructure, including towers and transmission lines, were destroyed rendering communication difficult and in many cases impossible. This region is also home to several underdeveloped remote villages and tiny communities where cellular penetration has not reached. Even the areas with coverage were badly damaged by the earthquakes and rehabilitation is still going on several years later. 

Responding to emergency situations is not only difficult but can sometimes be life-threatening. One of the most difficult aspects of rescue is ensuring efficient coordination among rescue team members on hostile terrain in an unfamiliar territory, such as crumbled buildings during an earthquake and buildings on fire. Rescue workers, separated by even a couple of hundred of meters of hostile terrain, can face problem in communicating. Moreover, communication in such scenarios is not necessarily restricted to radio communication but may also involve text messages, live video feeds, geographical positioning information, and triggered warning messages among others.  A dire need was felt for an appropriate wireless disaster communication system for these areas in responding to emergency situations for on-site coordination among the members of the rescue teams. In addition, such a system could also help in the post-disaster rehabilitation process spanning several months or even years. 

An effective and successful disaster communication system should have the following desirable characteristics:

\begin{itemize}

\item Should be cheap, rugged, easy to build, and easy to deploy (especially for rural deployments in developing countries);

\item Should have secure communication as well as protection against unauthorized access;

\item Should have inbuilt fault tolerance and resiliency through redundancy at all levels including multiple communication paths;

\item Should be flexible enough to effectively provide wireless coverage over diverse terrains; and

\item Should continue to work even when the the power grid or the conventional communication infrastructure (PSTN and cellular) is disrupted.

\end{itemize}

There are existing disaster communication systems that have been around for decades, however, each have some limitations. For instance, satellite phones provide universal coverage, however, during the earthquake of 2005 in North-Pakistan, the high cost of satellite phones prohibited their regular use. Radio-based communication has the limitations of range, bandwidth, and functionality and only provides poor quality audio broadcast. Several wireless base stations and various communication cables were destroyed during Hurricane Katrina and the remaining network was not able to support communication services to first responders \cite{Portmann2008wireless}. The legacy and tethered emergency response systems either are severely limited or completely destroyed in disaster struck regions due to their over dependence on the terrestrial infrastructure \cite{zhang2010wireless} and may not be able to support high bandwidth requirements for multimedia communication needed for disaster communication.

This paper presents the \textit{WiMesh system}, a secure and resilient wireless mesh network based disaster communication system designed, developed, and deployed at a conglomeration of remote and underdeveloped villages and tiny cities in the mountainous South-West region of Pakistan. The WiMesh system comprises of a two-tier architecture: the first tier consists of a portable, battery-powered wireless mesh network which is auto-configuring, self-healing, and rapidly-deployable; whereas the second tier consists of mobile handheld phones carried by rescue workers which connect automatically with the nearest mesh access point through Wi-Fi enabling the workers to communicate and coordinate with other team members. A prominent feature of the WiMesh system is that it is a complete product, and provides several useful features including audio communication (one-to-one as well as broadcast mode), SMS messaging facility, ability to see the live geographical positions of other members at all times, and multimedia streaming facility. These features, and the cost-effectiveness of the proposed approach, make the WiMesh system appealing for use in disaster areas or an alternative to traditional communication in remote areas. 


The major contributions of the paper are as follows:

\begin{enumerate}
\item WiMesh system details --- a secure and resilient disaster communication system based on multi-radio, multi-channel 802.11n wireless mesh nodes --- that provides audio, video, and text communication by exploiting the Wi-Fi capability of mobile phones in disaster-affected or remote and under-privileged areas where traditional communication and power infrastructure is non-existent or damaged;

\item The design and implementation details of the client and server applications, security considerations, routing protocols, and the resilient routing metric in light of the various design trade offs with respect to the constraints imposed by the infrastructure, equipment availability, and budget for developing countries. The feasibility and the performance of the WiMesh system is demonstrated using a thorough performance evaluation using both an indoor test bed and in a real-world outdoor setting; and

\item A discussion on common implementation challenges and pitfalls and sharing of general lessons that can guide researchers and practitioners in designing and developing future such systems. In addition,  the source code, design specifications, and implementation guidelines for the WiMesh system are also shared publicly to facilitate community development and knowledge sharing.

\end{enumerate}

The focus of this paper is not to propose yet another mesh modeling or routing protocol, but instead we believe that the state-of-the-art for mesh networks is quite mature and we focus instead on applying the research to solve a real-world problem while trying to balance several unique constraint imposed in the context of countries with very limited resources and semi-literate populace in remote villages who need a plug-and-play  solution. We do share some experimental results, but those serve more as validation of the concept rather than focus on providing deep insights on the functioning of the system. We target remote villages and rural areas with little to no communication and power-grid infrastructure to start with. Additionally, in contrast to the disaster communication systems designed for developed countries, financial constraints and scarcity of resources were the primary design considerations for WiMesh.  Moreover, the end-users of the WiMesh system comprise mostly of rural population with little to no knowledge or technical expertise. Thus, several of the design decisions were motivated by the cost constraints, availability of technology, and with a plug-and-play approach. The paper also provides rare insights into how  modern wireless communication technology is deployed and adopted by the rural population in Ziarat, a disaster-struck and harsh remote mountainous region in South-West Pakistan with its unique centuries old traditions and values. The paper also shares actual field results from the experiments conducted in Ziarat after WiMesh was deployed following the earth quake of 2008 in Ziarat with its unique geographic diversity.
\vspace{2mm}

The rest of the paper is organized as follows. Section II summarizes a review of the literature and the state of the art. The key differences are highlighted from existing works and the motivation for this work is discussed. Section III identifies key disaster communication system features and discusses design choices for these features.  Section IV provides the WiMesh system details including its overview, design considerations, architecture, and various components. Section V provides results from field trials for the performance evaluation of the WiMesh system along with challenges and problems faced during the development and deployment. The major insights and lessons learned are summarized in Section VI.  Section VII concludes the paper.

\section{Related Work}
Rising fatalities across the world, from natural and men-made disasters, have attracted significant research attention for disaster communication systems. Commonly deployed wireless disaster communication systems may be classified into three categories: cellular, hybrid, and mesh/ad hoc disaster communication systems. The cellular infrastructure has been widely used in several disaster communication systems \cite{shao2011rapid}\cite{zakia2016navigation} since it can provide significant coverage for areas with an existing and intact cellular infrastructure without the requirement of installing any additional hardware. However, these solutions are vulnerable as natural disasters often cause partial or complete damage to the telecommunication infrastructure. Moreover, the telecommunication infrastructure may not exist in remote areas of developing countries. Hybrid disaster communication systems \cite{martinez2016hybrid} leverage satellite connectivity in addition to the terrestrial infrastructure. As previously mentioned, the terrestrial infrastructure is vulnerable to disasters whereas the satellite subscription and equipment is costly, as was evidenced during the rescue operations in the 2005 Kashmir earthquake in the Northern Pakistan region \cite{KashmirEarthquake}\cite{KashmirEarthquake2} in which satellite phones could only sparingly be used. The mesh or ad hoc disaster communication systems \cite{chipara2012wiisard,george2010distressnet, minh2014fly, aiache2009chorist,aiache2005widens,del2009salice,de2005high} are based on leveraging wireless mesh and ad hoc networking to achieve rapid wireless multi-hop connectivity on disaster sites. Wireless mesh and ad hoc networks have several interesting features that make them an attractive option for disaster communication. Mesh networks offer wireless multihop connectivity, rapid deployment, minimal configuration, self-healing, and most importantly, these networks can often be built from off-the-shelf equipment. Due to these interesting features, mesh and ad hoc networking appears to be ideal candidates for disaster communication and, therefore, this section has focused on evaluating only mesh and ad hoc related network solutions.

\begin{table*}
\centering
\label{ComparativeAnalysis}
\caption{Disaster Communication Systems Comparison}
\scalebox{0.75}{
\begin{tabular}{|l|l|l|l|l|l|l|l|p{2.2cm}|}
\hline
\textbf{Features} & \begin{tabular}[c]{@{}l@{}}\textbf{WIISARD}\\ \end{tabular} & \begin{tabular}[c]{@{}l@{}}\textbf{DistressNet}\\ \end{tabular} & \begin{tabular}[c]{@{}l@{}}\textbf{OEMAN}\\ \end{tabular} & \begin{tabular}[c]{@{}l@{}}\textbf{CHORIST}\\ \end{tabular} & \begin{tabular}[c]{@{}l@{}}\textbf{WIDENS}\\ \end{tabular} & \begin{tabular}[c]{@{}l@{}}\textbf{SALICE}\\ \end{tabular} &
\begin{tabular}[c]{@{}l@{}}\textbf{SEDCOS}\\ \end{tabular} &
\textbf{\cellcolor{blue!25}\underline{WIMESH}}\\
\hline

\begin{tabular}[c]{@{}l@{}}Services \end{tabular} & \begin{tabular}[c]{@{}l@{}}First responder \\medical triage;\\Services for\\ victims \end{tabular} & \begin{tabular}[c]{@{}l@{}}Urban search \\and rescue;\\Internet \\connectivity for \\ social networking\\ between rescue \\workers \end{tabular} & \begin{tabular}[c]{@{}l@{}}Search and rescue;\\Internet connect-\\ivity for social\\ networking among \\rescue workers;\\Communication\\ among victims via\\ mobile devices \end{tabular} & \begin{tabular}[c]{@{}l@{}}Voice services;\\Applications for\\ public safety \end{tabular} & \begin{tabular}[c]{@{}l@{}}Reliable \\communication for \\real-time \\applications in\\disaster situations; \\Security and \\video surveillance \end{tabular} & \begin{tabular}[c]{@{}l@{}}Emergency \\communication, ~\\navigation \\and localization\\service\\for very large \\geographic areas\end{tabular} & \begin{tabular}[c]{@{}l@{}}Security, \\communication, ~\\authentication \\and confidentiality\end{tabular} & \vspace{-10mm} Secure and resilient services including multimedia (audio/video/SMS) and navigation and localization. \\ 
\hline
\begin{tabular}[c]{@{}l@{}}System \\Architecture \end{tabular} & \begin{tabular}[c]{@{}l@{}}Distributed, flexible \\and modular \\architecture \end{tabular} & Distributed & Fully centralized & \begin{tabular}[c]{@{}l@{}} Mainly centralized\\ with some\\distributed \\components \end{tabular} & \begin{tabular}[c]{@{}l@{}}Mainly centralized\\ with some\\distributed \\components\end{tabular} & \begin{tabular}[c]{@{}l@{}}Mainly centralized\\ with some\\distributed \\components\end{tabular} & \begin{tabular}[c]{@{}l@{}}Completely \\distributed\end{tabular}  & \vspace{-5mm} Decentralized distributed architecture.  \\ 
\hline
\begin{tabular}[c]{@{}l@{}}Wireless \\Network:\\ Structure, \\Technology,\\ and Coverage  \end{tabular} & \begin{tabular}[c]{@{}l@{}}802.11 for mesh\\ networking;\\Medium range \end{tabular} & \begin{tabular}[c]{@{}l@{}}802.11 for mesh\\ networking;\\802.15.4 for \\sensing;\\Short to medium \\range \end{tabular} & \begin{tabular}[c]{@{}l@{}}802.11n for Wi-Fi;\\ Medium range \end{tabular} & \begin{tabular}[c]{@{}l@{}}802.11n for edge\\ mobile first\\ responders;\\WiMax for \\vehicular backbone;\\ Long range \end{tabular} & \begin{tabular}[c]{@{}l@{}}OFDMA; \\MIMO; \\Long range \end{tabular} & \begin{tabular}[c]{@{}l@{}}IEEE 802.11;\\DVB-SH;\\ Very long range\end{tabular} & \begin{tabular}[c]{@{}l@{}}Bluetooth;\\ Very short range\end{tabular} & \vspace{-7mm} IEEE 802.11n mesh networking technology; Long range.  \\ 
\hline
Routing & \begin{tabular}[c]{@{}l@{}}Delay tolerant \\networking;\\Gossip based \\communication \\protocol (WCP);\\Hop count metric;\\AODV routing \end{tabular} & \begin{tabular}[c]{@{}l@{}}RPL routing \\protocol for low \\power and lossy\\ networks \end{tabular} & \begin{tabular}[c]{@{}l@{}}Tree-based \\unidirectional \\forwarding from\\ root to leaf nodes;\\No node-to-node\\ routing \end{tabular} & \begin{tabular}[c]{@{}l@{}}Layer 2-5 label\\ switching approach \end{tabular} & \begin{tabular}[c]{@{}l@{}}Secured sourced \\routing extension \\of OLSR protocol \end{tabular} & \begin{tabular}[c]{@{}l@{}}Extension of\\ODMRP routing\\protocol~\end{tabular} & \begin{tabular}[c]{@{}l@{}}Delay tolerant\\networking\\device-to-device~\end{tabular}& \vspace{-7mm} OLSR routing protocol with Expected Link Performance (ELP) metric. \\ 
\hline
\begin{tabular}[c]{@{}l@{}}Security \& \\Resilience\end{tabular} & \begin{tabular}[c]{@{}l@{}}Delay tolerant \\networking;\\Immune to comm.\\ delays;\\ No special\\ security measures\end{tabular} & \begin{tabular}[c]{@{}l@{}}Tolerant against \\delays and disruptions \\social networking\\ ensured at all times;\\No special security\\measures \end{tabular} & \begin{tabular}[c]{@{}l@{}}Limited fault \\tolerance due to \\centralized model;\\No special security\\measures \end{tabular} & \begin{tabular}[c]{@{}l@{}}Resilient backbone\\based on WiMax;\\High security \end{tabular} & \begin{tabular}[c]{@{}l@{}}Tolerance to \\link and node\\failures \\by using secure\\OLSR protocol;\\ Highly secure \\protocol \end{tabular} & \begin{tabular}[c]{@{}l@{}}Reliable centralized\\backbone;\\Highly secure\end{tabular} & \begin{tabular}[c]{@{}l@{}}Resilient to\\Denial of Service\\attacks through \\epidemic \\dissemination;\\ Specially design\\ for security\\ against DoS attacks\end{tabular} & \vspace{-10mm} Secure and Resilient. Uses (i) authentication; (ii) encryption (using WPA2-AES); (iii) resilient routing using ELP metric. \\ 
\hline
\begin{tabular}[c]{@{}l@{}}Power\\ Requirement \end{tabular} & Low & Low & Low & High & High & High & High & Low  \\ 
\hline
\begin{tabular}[c]{@{}l@{}}Software\\\& Operating\\ System \end{tabular} & \begin{tabular}[c]{@{}l@{}}Multiple \\OS support (Linux,\\ WinXP, Max OS) \end{tabular} & \begin{tabular}[c]{@{}l@{}}Multiple OS \\support (TinyOS, \\SOS, LiteOS, \\Mantis, Linux, \\Windows)\end{tabular} & Windows OS & \begin{tabular}[c]{@{}l@{}}Commercial \\proprietary software\end{tabular} & \begin{tabular}[c]{@{}l@{}}Embedded OS \\(PHY/MAC)\end{tabular} & Linux Ubuntu & \begin{tabular}[c]{@{}l@{}}Embedded OS \\(PHY/MAC)\end{tabular} & \vspace{-5mm} OpenWRT (Mesh Nodes); Android OS (WiMesh clients).  \\ 
\hline
\begin{tabular}[c]{@{}l@{}}Hardware\end{tabular} & \begin{tabular}[c]{@{}l@{}} Triage devices\\ (Nokia mobiles);\\ Mid-tier (Tablet PC);\\ Mesh boxes (Alix\\ x86);\\ RFID tags \end{tabular} & \begin{tabular}[c]{@{}l@{}}Mobile phones;\\Sensors;\\Wireless APs \end{tabular} & \begin{tabular}[c]{@{}l@{}} Laptops;\\Mobile phones \end{tabular} & \begin{tabular}[c]{@{}l@{}}Base station \\equipment \end{tabular} & PCMCIA Adapters & \begin{tabular}[c]{@{}l@{}}Software-defined\\radios, laptops,\\ mobile base \\stations\end{tabular} & \begin{tabular}[c]{@{}l@{}}Software-defined\\radios, laptops,\\ mobile base \\stations\end{tabular} & \vspace{-7mm} Ubiquiti Picostation and the Nanostation devices.  \\ 
\hline
\begin{tabular}[c]{@{}l@{}}Size and \\Portability \end{tabular} & \begin{tabular}[c]{@{}l@{}}Small size;\\Light;\\Highly portable \end{tabular} & \begin{tabular}[c]{@{}l@{}}Small size;\\Light;\\ Highly portable \end{tabular} & \begin{tabular}[c]{@{}l@{}}Small size;\\Light;\\Highly portable \end{tabular} & \begin{tabular}[c]{@{}l@{}}Large size;\\Heavy;\\Portability difficult -\\requires vehicles \end{tabular} & \begin{tabular}[c]{@{}l@{}}Small size;\\ Light;\\Highly portable \end{tabular} & \begin{tabular}[c]{@{}l@{}}Not portable,\\ requires \\infrastructure\end{tabular} & \begin{tabular}[c]{@{}l@{}}Not portable,\\ requires \\infrastructure\end{tabular} & Highly Portable.  \\ 
\hline
\begin{tabular}[c]{@{}l@{}}Installation \\Complexity\end{tabular} & Low & Low & Low~ & High & Medium & High~ & High& \begin{tabular}[c]{@{}l@{}} Low \end{tabular}  \\ 
\hline
Cost & Low & Low & Low & High & Low & High & High& \begin{tabular}[c]{@{}l@{}} Low \end{tabular}  \\ 
\hline
Availability & \begin{tabular}[c]{@{}l@{}}Commercial Off-the-\\Shelf (COTS);\\Digital tags may \\be difficult to \\acquire in develop-\\ing countries \end{tabular} & COTS & COTS & \begin{tabular}[c]{@{}l@{}}Required hardware \\for base stations \\and associated\\ network may be \\difficult to acquire\\ at remote sites \end{tabular} & \begin{tabular}[c]{@{}l@{}}Custom hardware\\and software\end{tabular} & \begin{tabular}[c]{@{}l@{}}Custom \\hardware\end{tabular} & \begin{tabular}[c]{@{}l@{}}Custom \\hardware\end{tabular}& \vspace{-7mm}  Relatively cheap and easily available equipment is used. \\ 
\hline
\begin{tabular}[c]{@{}l@{}}Major\\ Contributions \end{tabular} & \begin{tabular}[c]{@{}l@{}} Mitigates network \\partitions due to\\ mobility of rescue\\ workers using DTN;\\Characterization of\\ link quality and \\human mobility\\ patterns during \\medical triage \\among mobile first\\ responders \end{tabular} & \begin{tabular}[c]{@{}l@{}}Offers file storage;\\Provides social \\networking despite\\ network delays\\ and disruptions;\\Mobile vibration \\sensing application\\ detecting victims \\under rubble with \\improved accuracy \end{tabular} & \begin{tabular}[c]{@{}l@{}}Converts \\commodity mobile \\devices into Virtual\\ Access Points \\(VAPs);\\Supports extensions\\ of Internet \\connectivity to\\ disaster sites \end{tabular} & \begin{tabular}[c]{@{}l@{}}Provides video \\surveillance and\\ voice communica-\\tion for disaster \\sites by providing \\a WiMax VANETs \\for vehicles to \\which mobile first\\ responders connect\\ through Wi-Fi \end{tabular} & \begin{tabular}[c]{@{}l@{}}Cross-layered \\approach for high\\ data rate service;\\Terminode concept\\ where each node \\performs the \\function of cluster\\ head, relay, router, \\or gateway;\\Secured source \\routing \end{tabular} & \begin{tabular}[c]{@{}l@{}}Platform based \\on SDR \\technology for \\emergency\\services; \\Integration of\\different \\components and \\technologies.\end{tabular}& \begin{tabular}[c]{@{}l@{}}Platform based \\on SDR \\technology for \\emergency\\services; \\Integration of\\different \\components and \\technologies.\end{tabular} & \vspace{-15mm} A resilient WMN based disaster communication system for underdeveloped/rural areas using low-cost, highly available, portable equipment.  \\ 
\hline
Limitations & \begin{tabular}[c]{@{}l@{}}Sub optimal \\routing metric \\hop count\end{tabular} & \begin{tabular}[c]{@{}l@{}}Limited experimen-\\tal evaluations \\performed indoors \\only using a piece-\\wise approach e.g. \\mobility algorithm, \\sensing, etc.\end{tabular} & \begin{tabular}[c]{@{}l@{}}Internet extensions \\only for rescue staff;\\Limited indoor \\evaluation only; \\Does not support \\peer-to-peer\\communication,\\only root-to-node;\\Does not perform \\load balancing, uses\\a single AP\end{tabular} & \begin{tabular}[c]{@{}l@{}}Expensive;\\Uses licensed\\frequency bands\end{tabular} & \begin{tabular}[c]{@{}l@{}}Uses licensed\\frequency bands\end{tabular} & \begin{tabular}[c]{@{}l@{}}Platform offers \\limited through-\\put due to the \\limitations of the\\ USRP in NLOS \\conditions\end{tabular} & \begin{tabular}[c]{@{}l@{}}Platform offers \\limited through-\\put due to the \\limitations of the\\ USRP in NLOS \\conditions\end{tabular}& \vspace{-7mm} Due to its specific design for disaster communication scenarios, it may not work well for other high-performance requirements.  \\ 
\hline
\end{tabular}}
\label{ComparativeAnalysis}
\end{table*}

\textit{WIISARD} \cite{chipara2012wiisard} is an emergency response system for providing medical triage services by mobile first responders. The system initially used Ad hoc On-Demand Distance Vector (AODV), but later switched to a gossip-based communication protocol for delay tolerant networking. The main contribution of this research work is the characterization of wireless links and human mobility patterns during medial triage among mobile first responders. The research work presented in \cite{chipara2012wiisard} showed that network partition may occur and, therefore, proposed a gossip protocol for delay tolerant communication in the mesh IEEE SYSTEMS 3 backbone. One of the limitations of WIISARD system is that it uses a rudimentary routing metric of hop count that has been shown to perform sub-optimally \cite{de2005high}.

\textit{DistressNet} \cite{george2010distressnet} is an emergency response system for Urban Search and Rescue (USR) in disaster struck areas. Low-powered COTS (Commercial Off-the-Shelf) devices, such as smartphones and sensors, are used to assist in disaster communication. DistressNet uses IPv6 based routing protocol for low-power and lossy networks (RPL). The major contributions include file storage and social networking services offered over a delay tolerant network. DistressNet also supports a vibration sensing application that helps detect victims trapped under rubble. The main limitation of this solution is that the performance evaluation is performed on a piecemeal basis and lacks results for a holistic and comprehensive performance evaluation.

\textit{OEMAN} (On-the-fly Establishment of Multihop wireless Access Networks) \cite{minh2014fly} is a novel approach for disaster communications. This system leverages the availability of commodity mobile devices at disaster sites, turning them into virtual apps to extend Internet connectivity for disaster sites. There are several limitations including implementation and evaluation in a limited indoor building setting, support for only root top AP route discovery without any support for node-to-node communication, and scalability problems.

\textit{CHORIST} (Communications for enHanced envirOnmental RISk management and citizens safeTy) \cite{aiache2009chorist} system forms a WiMAX based mesh Vehicular Ad hoc NETwork (VANET) between rescue vehicles. Mobile first responders who are located at the edge of the network can connect to this backbone through Wi-Fi. The system offers voice services and video surveillance for disaster sites. A layer 2.5 label switching approach is used for data communication. The main limitation of this solution is that it requires expensive and time consuming WiMAX base stations infrastructure setup for the wireless backbone and the availability of abundant vehicles. Moreover, the cost of the WiMAX licensed band may also limit the utility of this approach. 

\textit{WIDENS} (WIreless DEployable Network System) \cite{aiache2005widens} uses a reliable communication system for real-time applications in disaster situations. This system adopts a cross-layered, cluster-based approach to provide high speed communication through hotspots and exploits the high bandwidth and wireless range of the MIMO (Multiple Input Multiple Output) at the physical layer. This system introduces the concept of ``Terminodes'' in which each node in the network can perform the functions of a cluster-head, relay, router, or gateway, thereby, offering flexibility. The routing protocol used by the proposed solution is based on secure source routing, which addresses some security limitations in other similar systems. The main limitation of this solution, however, is that it uses licensed frequency bands, which may limit its use or may introduce excessive costs in some countries.

\textit{SALICE} (Satellite-Assisted Localization and Communication system for Emergency services) system \cite{del2009salice} uses a hybrid disaster communication comprising of ground infrastructure as well as satellite connectivity. This work employs several different concepts including satellite-assisted localization, Software-Defined Radios (SDR), Cognitive Radios (CR), and satellite-assisted enhanced reachability for rescue workers. Due to these powerful and diverse technologies, the SALICE system offers comprehensive functionality with flexibility and heterogeneity for emergency situations. This research work highlights the benefits of the SDR technology for addressing the dynamic nature of emergency situations allowing new communication links or adoption of modulation schemes based on the ground situation. The main limitation of this solution is its lack of cost-effectiveness due to a comprehensive infrastructure. 

\textit{SEDCOS} (A Secure Device-to-Device Communication System for Disaster Scenarios) \cite{SEDCOS} has been especially designed with security considerations. It adopts an adversary based model and employs secure key management and resilient communication to counter denial of service (DoS) attacks. It does not discuss classical disaster communication services such as audio message and voice calls, rather contributes towards modern disaster communication systems.

Table \ref{ComparativeAnalysis} provides a comparative analysis of the disaster communication systems presented in this section. 

A number of mesh network based disaster communication systems are evaluated above.  While these systems have positively contributed in the overall mesh network based disaster communication research, all of these systems have limitations as was highlighted in the above evaluation. The proposed WiMesh system is expected to provide similar benefits while attempting to overcome some of these limitations.

\section{Design Considerations for Components of Disaster Communication Systems}
\label{Design Considerations}
Designing efficient and effective disaster communication systems involves several diverse components of the system working seamlessly to provide the required performance. Towards this end, this section provides details on various key components of disaster communication systems and discusses possible design choices along with trade-offs. 

\vspace{2mm}
\subsection{Services}
\label{services}
Disaster communication systems are expected to provide following services:
\begin{enumerate}
\item User registration and authentication - User registration is required at the server whereas authentication is needed both at the server and the client level;
\item Audio call - Audio calls can be either one-to-one or one-to-many (broadcast). One-to-one audio calls may be between admin and users or among users;
\item SMS - SMS may also be either one-to-one or one-to-many (broadcast). Like audio call, one-to-one SMS may be between admin and users or among users;
\item File transfer - File transfer may also be either one-to-one or one-to-many. Like audio call, one-to-one file transfer may be between admin and users or among users;
\item Live video streaming - Live video streaming functionality may consist of following types: sending live video stream to client, sending live video stream  to  server,  receiving  live  video  stream  by  client,  and receiving  live  video  stream  by  server.  All  live  video  streams are usually one-to-one between two users;  
\item Navigation - The  navigation  functionality  allows  users  to  view  the  geographical  location  of  other  peers/users  on  a  map with some necessary information. User can also view his own position on the same map relative to the other peers/users; and
\item Network performance monitoring - Disaster communication systems need capability for monitoring, logging, and viewing user and network statistics to ensure QoS for the coordinated relief efforts among geographically dispersed clients in adverse situations.
\end{enumerate}

\vspace{2mm}
\subsection{System Architecture}
The system architecture plays an important role in the overall performance of any disaster communication system. The specific choice of architecture may be influenced by several external factors including terrain, weather, existing infrastructure, and cost. 

Broadly, there can be three categories for system architecture: centralized, distributed, and hybrid. In general, the loosely coupled distributed architecture is the best choice for disaster communication due to a number of reasons including flexibility, fault tolerance, autonomy, and scalability. Moreover, logistic reasons, such as difficult terrains, also dictate that autonomous distributed components are more likely to survive the stringent requirements of disaster communication. However, the distributed architecture do introduce some security concerns due to lack of control. Popular solutions in the distributed architecture category include WIISARD\cite{chipara2012wiisard} and DistressNet \cite{george2010distressnet}. At the other end of the spectrum, centralized solutions, such as OEMAN\cite{minh2014fly}, even with some limitations, offer certain appealing features such as ease of use, control, and security. Some solutions \cite{aiache2009chorist,aiache2005widens,del2009salice} offer a compromise by combining the strengths of both these approaches. In general, these solutions have a fixed backbone (e.g., WiMAX or cellular) and a mobile mesh based last-mile. These solutions do offer a trade-off, however, the centralized architecture imposes the same limitations of scalability, flexibility, and cost.

\vspace{2mm}
\subsection{Wireless Network: Structure, Technology, and Coverage}
The wireless network is perhaps the most fundamental component of the disaster communication system and consists of several key aspects including its structure, technology, and coverage. The structure of a wireless network can either be centralized, distributed or hybrid. Distributed structures, such as ad hoc and mesh networks \cite{chipara2012wiisard,george2010distressnet,minh2014fly}, offer the inherent advantages of multi-hop networking, while a hybrid approach \cite{aiache2009chorist,aiache2005widens,del2009salice} combines some form of a centralized backbone network (e.g. satellite or cellular) along with a mesh component. 

Another important decision in the design of the WiMesh system is the choice of the wireless technology. Technologies like WiMAX or 4G offer certain advantages, however, have the limitation of infrastructure and licensing costs. Similarly, different technologies offer a trade-off in terms of wireless coverage as well. For instance, installing WiMAX or cellular towers can significantly increase the wireless reach of the network, however, have cost and logistic difficulty offsets. Similarly, off-the-shelf technologies like 802.11 are free and require minimal setup, however, have coverage limitations. In Wi-Fi technologies, the IEEE 802.11n standard uses Multiple-Input and Multiple-Output (MIMO), and provides significantly better performance by exploiting spatial and temporal diversity through multi-path propagation and offers impressive data rates of up to 600 Mbps. Therefore, IEEE 802.11n based radios are a natural choice for the mesh backhaul as well as offering access to clients. 

\vspace{2mm}
\subsection{Routing}
Routing forms the backbone of any disaster communication system and the routing protocol and metric used play a decisive role in the overall system performance. Towards this end, there are choices of several routing protocols and metrics including proactive, reactive, and hybrid routing solutions. Some completely centralized solutions employ static routing, but by and large, most disaster communication systems need a strong proactive or dynamic routing protocol. Popular choices in this category include OLSR and DSR routing protocols. Proactive protocols such as OLSR offer quick routes, however, create messaging overhead that can be burdensome if the equipment is battery powered. On the other hand, the reactive routing approaches do not provide quick routes, however, can offer overhead savings. Another important criteria is that of the routing metric as the metric needs to be dynamic and integrates fault-tolerance and adaptability to changing network conditions. This is especially important in disaster situations since the network may experience failures and faults due to users moving out or range, environmental factors, and power draining. State-of-the-art metrics such as \cite{ashraf2013route} can offer efficient routing metric solutions which cater to changing link qualities, bandwidth, and node failures.

\vspace{2mm}
\subsection{Security and Resilience}
Disaster communication system, by virtue of its application, may have several diverse clients connected to its network. These clients may consist of mobile devices such as smart phones, laptops, tablets, and other communication gadgets of either official rescue workers or personal users. These devices may be used for rescue communications and data sharing or people requesting for aid or other emergency services. Due to this wireless, cooperative, and decentralized nature, these systems are vulnerable to various security attacks such as relief communication disruption by terrorists by injecting false messages, denial-of-service (DoS), and unintentional clogging of the network by spamming from people in panic \cite{SEDCOS}. Moreover, malicious attacks could damage strategic nodes or links in the network and, therefore, the underlying routing protocols and metrics should be designed to introduce resilience against these types of attacks. Hence, effective disaster systems must provide user authentication, message integrity, secure routing, confidential end-to-end communication and resilient routing.

\vspace{2mm}
\subsection{Power Requirement}
Since disaster communication systems are envisaged to be deployed in rural and disaster-struck areas, an implicit assumption is that electrical supply will be limited or unavailable. To solve this problem, renewable energy systems, such as solar power, may be considered to power the nodes. This implies that the designed system must be energy-efficient. Moreover, the software on the mobile phones must be designed to minimize the energy drain. 

\vspace{2mm}
\subsection{Software and Operating System}
There are three main parts of disaster communication software system including the server, the clients, and the wireless back haul network OS/firmware.

\subsubsection{OS for Wireless Network}
Proprietary OS and firmware provide a user-friendly graphical user interface.  However, there is little control over different network components such as the routing protocol, the routing metric, and dynamic rate adaptation. Therefore, disaster communication systems should consider the adaptability and flexibility of open-source software.

\textit{OpenEmbedded} \cite{OpenEmbedded} is an open-source cross-compilation environment for building Linux distributions for embedded applications. It offers support for multiple hardware, customization, and rapid development. A set of rules and instructions, called a ``recipe,'' is used to generate architecture specific applications and packages. However, OpenEmbedded is difficult to use due to its complexity.

\textit{Linux From Scratch (LFS)} \cite{LinuxFromScratch} offers a powerful paradigm for building custom Linux distributions from the source. The result is a compact, flexible, and secure system with a better understanding of the internal working of the OS. However, building LFS requires significant Linux proficiency to sort out the configuration dependencies and compilation errors.  

\textit{DD-WRT} \cite{dd-wrt} is a Linux based firmware for wireless routers and access points. DD-WRT provides a custom firmware option with several features and functionalities such as access control, bandwidth monitoring, quality of service, Dynamic DNS, universal plug-and-play, and many others. 

\textit{OpenWRT} \cite{OpenWrt} is an open-source Linux for embedded devices for routing network traffic. OpenWRT uses a smaller µClibc C library to reduce the memory footprint, which makes it possible to run on various types of devices, such as home routers, residential gateways, smartphones, pocket computers, and laptops. OpenWRT offers a built-in package manager that allows the installation of packages from a comprehensive software repository.  OpenWRT can be used as an SSH server, a Bit-Torrent client, and VPN. OpenWRT offers significant flexibility and has an extensive support base.

\vspace{2mm}
\subsubsection{Software for Clients}
There are several mobile application development platforms available for the client OS. The Mobile Information Device Profile (MIDP) \cite{midp} is the toolkit from Sun for creating mobile applications and is considered quite easy to learn and use. The Symbian OS \cite{SymbianOS} is another popular platform for mobile phones and is used by Nokia in their mobile phones. Symbian OS is significantly more complex compared to MIDP. The iPhone OS is the operating system used in Apples iPhone and iPod touch. iOS is designed to be extremely efficient on mobile devices, using less than half a gigabyte of memory \cite{Appleios}. The BlackBerry platform \cite{BlackberryOS} uses Java for application development and runs applications in protected run-time environment. It has several libraries with sound inter-operability. BlackBerry has acceptable level of robustness and reliability, however, the BlackBerry platform does not allow programmers controlling the GPS and scanning of cell towers or Wi-Fi access points (APs) programmatically.  Moreover, it is not very memory efficient.  The Pocket PC/Windows Mobile OS is a mobile version of the desktop Windows OS. It has been designed mainly for Pocket PC and Smartphone devices and as such limits the third party development opportunities.

\vspace{2mm}
\subsubsection{Software for Server}
\label{sw_server}
Every disaster communication system requires the usage of VoIP technology to support messaging and audio communication. While there are several dozen VoIP servers available, to limit the scope of discussion, this paper only considers the two major VoIP servers, Asterisk \cite{asterisk} and OpenSIPS \cite{opensips}. Asterisk is a software implementation of the telephone Private Branch Exchange (PBX). Asterisk enables telephone calls between users connected to Asterisk as well as to other telephone services such as the PSTN and VoIP services. Asterisk is compatible with several operating systems including NetBSD, OpenBSD, FreeBSD, macOS, and Solaris. Asterisk supports several standard VoIP protocols, including the Media Gateway Control Protocol (MGCP), Session Initiation Protocol (SIP), and H.323. The OpenSIPS is basically a multifunctional, multi-purpose signaling SIP server for VoIP that can handle voice, text, and video communication. OpenSIPS can handle thousands of calls simultaneously. The fundamental difference between the two VoIP servers is that Asterisk is basically a media server, whereas OpenSIPS is essentially a SIP proxy server. While Asterisk has a better support for multimedia communication, the robustness, flexibility, adherence to SIP standards, makes OpenSIPS a suitable choice for the disaster communication systems.

\vspace{2mm}
\subsection{Hardware}
The major choice of hardware is in the realm of the infrastructure of the wireless network side including Access Points (APs) and antennas among others. The backbone for disaster communication can be created ranging from relatively simple off-the-shelf wireless access points to powerful network boards like Gateworks Cambria boards \cite{GateworksCambria}. The higher end network boards are more powerful, rugged, and typically have multi-radio support, however, are more expensive and complex. Here, it is important to realize that there are important aspects including the cost, energy consumption, availability and robustness of the hardware which may play an important role in decision making. It is hard to deploy a solution which addresses problems of underdeveloped countries but requires importing expensive equipment that is not readily available. Moreover, the mesh nodes should also be easily portable and rugged since the equipment may need to be rapidly deployed over difficult geographical terrain. The mesh nodes should also be energy-efficient and would need to be solar-powered out in the field due to the likely destruction or non-availability of power grid in the disaster areas.

\vspace{2mm}
\subsection{Size and Portability}
Since disaster communication systems may be used over unfamiliar and uneven terrain, under hostile circumstances, and in inclement weather conditions, these systems should be easy to assemble, disassemble, carry, and move around. Often vehicles are not available or offer limited maneuverability. Hence, such systems need to be designed considering smaller size, lighter weight, and high portability.

\vspace{2mm}
\subsection{Installation Complexity}
A major factor which can dictate the eventual adoption and success of disaster communication systems is the installation complexity of the complete system. The installation complexity may include the physical deployment as well as configuring the server and client software. For disaster-struck, underdeveloped areas, it may be difficult to have highly skilled people. It would be pointless to add to the list of problems the complexity of deploying and configuring the disaster communication systems. A plug-and-play approach for all components of the system should be followed and extensive use of automated scripts for accomplishing repetitive tasks including bootstrapping common settings is recommended.

\vspace{2mm}
\subsection{Cost}
Since these disaster communication systems are expected to be used in developing and underdeveloped countries, the cost is an important factor to be considered in the design of such systems. Cost factor may consist of hardware equipment cost, software cost, maintenance cost, and skills cost. The choice of the hardware equipment should be as such that the cost is minimal and affordable in the developing countries.  The Software cost may be minimized or eliminated by selecting open source and freeware software. The design of the system should be flexible and adaptable so that maintenance cost is minimal. The system should be easy to setup and use to eliminate need of highly skillful operators.

\vspace{2mm}
\subsection{Availability}
Availability of the equipment is an important consideration especially for the developing countries. Often there is a choice of employing state-of-the-art equipment which is highly efficient and effective, however, their availability may not be as widespread. For instance, the Avila Gateworks board \cite{avila} is considered a  powerful solution for outdoor wireless networking, however, is not easily available in developing countries.

\section{The WiMesh System}
This section presents architectural, component design, and implementation details of the WiMesh system. In addition, design choices for the WiMesh system are presented for the system features identified in section \ref{Design Considerations}.
\subsection{Design Choices for WiMesh}
\subsubsection{Services}
Based on the key services identified in section \ref{services} for disaster communication systems, following services were selected to be incorporated in the WiMesh system: audio calls in both unicast and broadcast modes; multimedia data exchanges such as images, video, data files, and SMS in both unicast and broadcast modes; sending and receiving live video streaming to/from clients and server; and geographical position viewing of the WiMesh users. Moreover, the WiMesh controller provides additional services such as remotely monitoring, configuring, and controlling the entire network even at the wireless interface level. Additionally, the server also provides user registration and authentication services.

\subsubsection{System Architecture}
For the WiMesh system, a decentralized architecture was selected to avoid the inherent limitations of centralized networks. The decentralized architecture was combined with the power of mesh networking, exploiting its traditional strength areas such as multiple available routes, adaptability of changing requirements, and flexibility. Thus, the WiMesh system has been designed to enable loose coupling among autonomously working components in order to avoid a purely centralized approach. The server does introduce some centralized components, however, the routing in the mesh backbone itself is highly flexible, independent and adaptable to changing environment and network conditions.

\subsubsection{Wireless Technology and Coverage}
For the wireless technology, the off-the-shelf license free IEEE 802.11n technology was selected for the WiMesh system. There is the trade-off of limited wireless coverage at hand, however, by using 802.11n technology and powerful outdoor multiple sector antennas, the wireless range issues can be somewhat mitigated.

\subsubsection{Routing}
The OLSR protocol was selected for routing in the WiMesh network since it is a proactive protocol allowing for real-time routing decisions. Moreover, there are several implementations available for this protocol. For the routing metric, the Expected Link Performance (ELP) routing metric was integrated since it provides state-of-the-art performance by incorporating aspects such as resilience, link qualities, interference,load, and link asymmetries. 

\subsubsection{Security and Resilience}
The WiMesh server security and resilience was designed from three aspects. First and foremost, the WiMesh system registers and authenticates users (Authentication API module) in order to ensure that only authorized users access the services (there is a special emergency mode available which allows all users to register and connect without authentication). Second, the wireless communication is completely encrypted using WPA2-AES encryption for the wireless state-of-the-art encryption standards. Third, the ELP metric introduces resilience in the overall routing approach in the sense that nodes and link failures would be mitigated through the use of efficient and dynamic estimation of route qualities.

\subsubsection{Power Requirement}
The WiMesh system is designed to function in remote areas through the use of batteries and solar panels. This has been demonstrated through several experiments that none of the equipment is particularly power hungry including the server, controllers, and the mesh nodes and the system can function perfectly off-the-grid. Additionally, the WiMesh system conserves energy by minimizing the usage of the GPS module through periodic sleeping.

\subsubsection{Software and Operating System}
For the wireless back-haul network, the obvious choice for the WiMesh system was an open-source Linux distribution like OpenWRT or DD-WRT because these OS/firmware are easy to control, modify, and extend. Significant changes can be made to the protocol stack for designing efficient protocols and integrating customized code and scripts. Therefore, OpenWRT was used as the Linux distribution for the mesh nodes.

For the server software, Asterisk has a better support for multimedia communication, robustness, flexibility, and adherence to SIP standards. However, it seemed that the speed and call capacity of the OpenSIPS server, along with its tremendous scalability capacity, makes it an ideal choice for the WiMesh server. 

The Android OS \cite{AndroidOS} open-source development platform was selected for the WiMesh mobile clients. Android OS offers component-based architecture, built-in APIs, automatic lifecycle management, and high portability.

\subsubsection{Hardware}
The principal hardware considerations were cost effectiveness, off-the-shelf availability, and open-source programmability for full control over device functionality. Hence, the hardware choice for this project was the Ubiquiti Picostation and the Nanostation devices \cite{ubiquiti}.

\subsubsection{Size and Portability}
Highly portable equipment was selected, e.g., extensible poles that can be quickly extended or collapsed. The wireless APs and other devices were also lightweight and portable. No heavy installations were required, making the WiMesh system a highly portable solution. 

\subsubsection{Installation Complexity}
Extensive use of bootstrapped automated scripts  were used to ensure common settings and easy deployment. The OpenSIPS server and the server software was almost plug-and-play. The final product was a professionally assembled unit with LEDs and simply required turning it on. For the clients, an android application was developed which could be easily installed.  

\subsubsection{Cost}
Instead of importing expensive wireless equipment, the project instead relied on cost-effective equipment locally available. Redundancy was introduced to offset some of the quality trade-offs in the solution. For example, instead of power Avila Gateworks boards with multiple Wi-Fi cards support \cite{avila}, multiple Ubiquiti wireless APs \cite{ubiquiti} with multiple sector antennas were used per mesh node.

\subsubsection{Availability}
Avila Gateworks boards \cite{avila} were not available locally and had to be imported, which were costly and would pose problems for additional parts replacement problems down the road. Therefore, it was decided to use relatively cheap and easily available equipment, e.g., Ubiquiti Nanostations and Picostations \cite{ubiquiti} and antennas that were readily available. The trade-off was quality and performance, but redundancy (multiple APs per mesh node and more dense deployment) was used to overcome these limitations.

\begin{figure}[t]
\centering
\includegraphics[width=8cm, height=5.2cm]{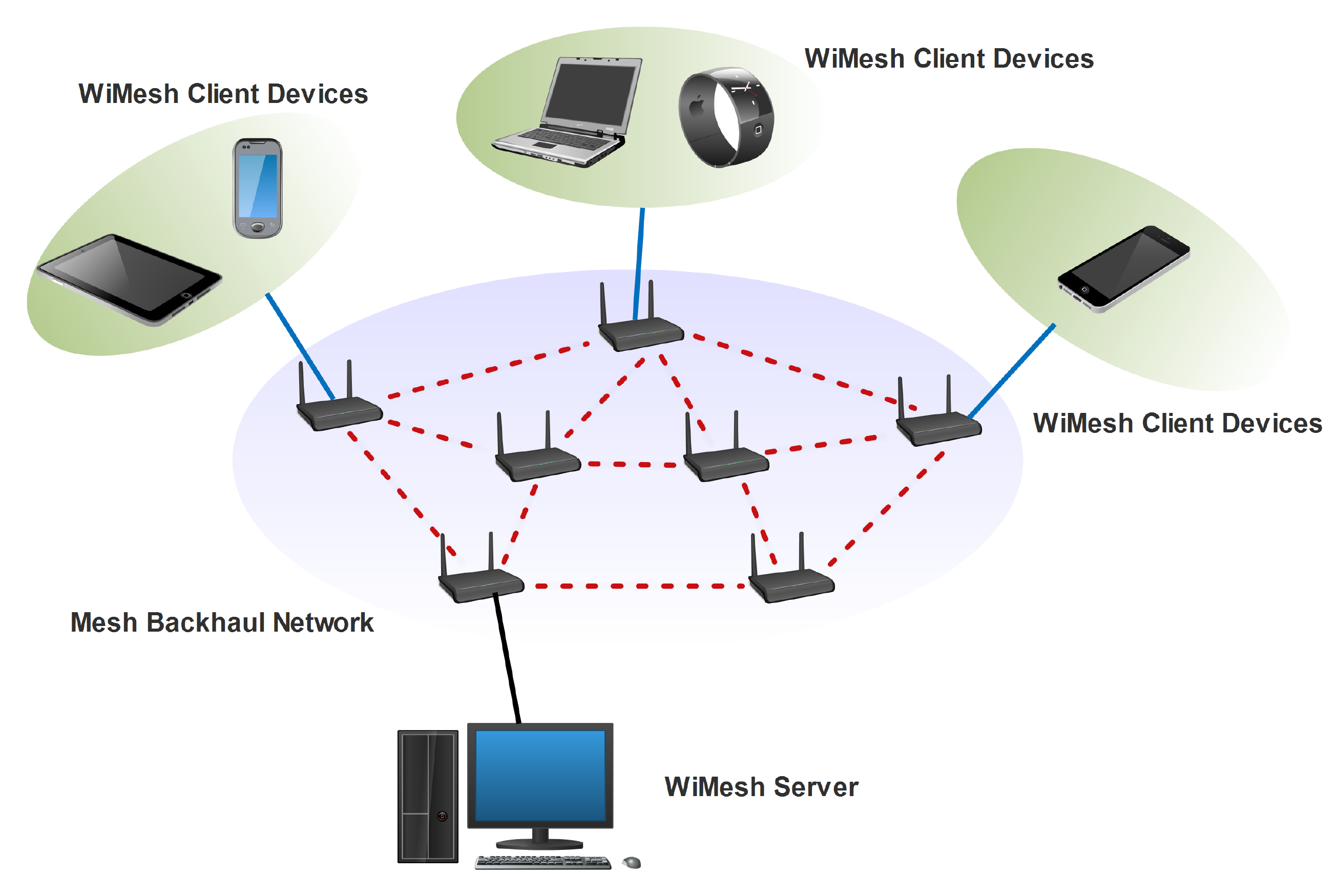}
\captionsetup{singlelinecheck=true} 
\caption{\label{Arch} WiMesh Architecture}
\end{figure}

\subsection{WiMesh Architecture and Overview}
Figure \ref{Arch} shows the WiMesh system architecture with its three main components: server module, client module, and the backhaul mesh network. The WiMesh server module is responsible for the overall management of the WiMesh system. The server module provides mechanisms for all communications between clients and rest of the WiMesh network including audio, video, messaging, file exchange, and live geographical positioning of clients. The server also handles user registration and authentication. In addition, the server module monitors, collects, and logs user data and the performance statistics of the WiMesh backhaul network. The server module is typically hosted on a desktop machine. The WiMesh client module is installed on mobile devices carried by the users and enables multimedia communication and live geographical positioning with other clients and the WiMesh server. The WiMesh backhaul network is a multi-radio, multi-channel wireless (Wi-Fi) backhaul mesh network that forms the core of the WiMesh network.  The backhaul mesh network creates a wireless multi-hop backbone infrastructure enabling mobile users, spread over large geographical regions, to communicate with each other.  The WiMesh backhaul network uses the Optimized Link State Routing (OLSR) protocol, which is integrated with a custom WiMesh routing metric to provide efficient custom routing for performance improvement.

\subsection{WiMesh Server}
The WiMesh server is an important part of the WiMesh system, which handles communication among clients as well as the overall system including the mesh backhaul network. The core functionalities offered by the server include:

\begin{itemize}
\item Overall management of the system;
\item Audio calls in broadcast mode with WiMesh clients;
\item Multimedia exchanges (image, video, data files, SMS) with clients;
\item Geographical position viewing of the WiMesh users;
\item Registration and authentication of WiMesh clients; and
\item User data and mesh network statistics monitoring and logging via WiMesh Controller.
\end{itemize}

Some of the functionalities such as audio calls, multimedia exchanges, and geographical positions viewing require interactions with the clients and, hence, are repeated in the WiMesh client modules as well.  The client-side part of these functionalities will be discussed in the client module section.

\begin{figure}[t]
\centering
\includegraphics[width=8cm, height=9cm]{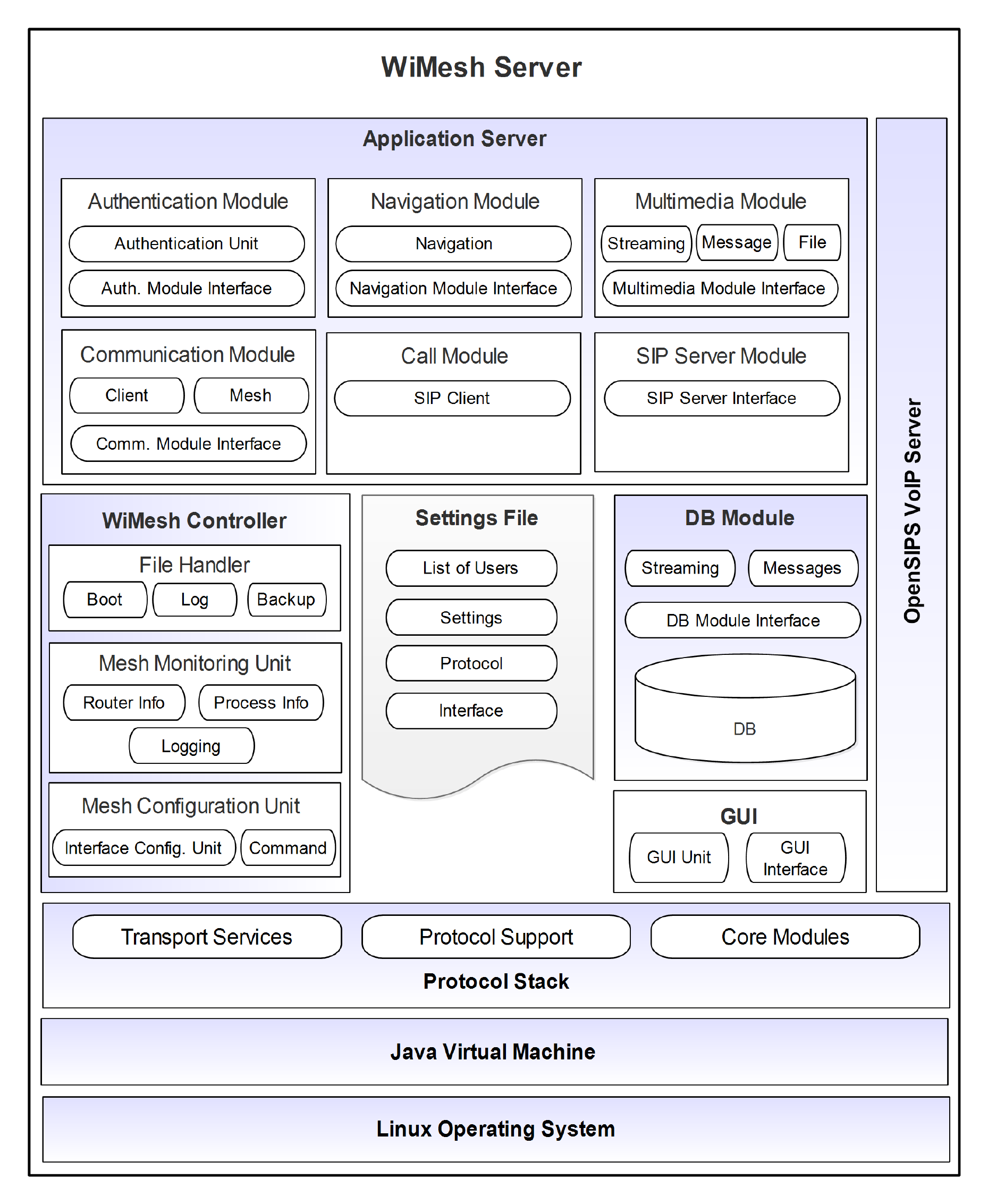}
\captionsetup{singlelinecheck=true} 
\caption{\label{ServerArch} WiMesh Server Architecture}
\end{figure}

Figure \ref{ServerArch} shows the main components of the WiMesh server module.  The server module hosts an application server, an OpenSIPS VoIP server, and a WiMesh controller.  Both the application and the OpenSIPS VoIP servers are hosted on the same machine with an underlying protocol stack consisting of various transport services.  The server module is programmed in the Java programming language running on top of a Java Virtual Machine (JVM) in a Linux operating system environment.  In addition to these main components, the server module also consists of a database module, graphical user interface (GUI) components, and various configurations files.

\vspace{2mm}
\subsubsection{Application Server}
The application server comprises of the different service APIs.  These APIs consist of VoIP, text, file, video and the location/navigation API.  In addition, the application server also contains a communication module that comprises of Mesh APIs for interfacing with the mesh network and managing the overall operation of the server. Finally, an authentication module provides client registration and authentication functionality.

\textit{Call APIs}: The call APIs handle audio communication between the server and the OpenSIPS server. The audio communication supported by the application server is of two types: the real-time one-to-one VoIP call mode and the audio broadcast mode. The real-time one-to-one VoIP call is only supported on the WiMesh clients, such as mobile phones, whereas the audio broadcast mode is used on the WiMesh server by the administrator to send out a broadcast audio to all users through a radio-like functionality.  For the broadcast audio messages, the administrator on the WiMesh server presses and holds a ``record'' button to record audio messages of up to two minutes. The audio message is encoded using Adaptive Multi-Rate audio codec and transmitted using HTTP Live Streaming (HLS) to clients. 

The VoIP APIs are implemented on top of open-source Mobicents Application Server platform, which is a high-performance, scalable, and fault tolerant service level execution environment. The VoIP API implementation focuses on enabling the establishment and maintenance of audio communication between mobile users. From the implementation point of view, the VoIP API uses the Mobicents Application Server’s SIP module and the SIP protocol for voice communication. A sub-module of the VoIP API allows the command and control station to communicate with the mobile users. 

\textit{Multimedia APIs}: The multimedia APIs allow sending a file or a message to users from the WiMesh server. Using these APIs, the administrator can send a message or a file to a specific user or broadcast to all users. In addition, the multimedia APIs allow streaming live video from WiMesh server to the clients. The purpose of this functionality is to facilitate monitoring of real-time ground situation.

For the file and text APIs, as soon as a text message is received from a user destined for another user, the server checks if the destination is available. If so, the message is sent and a copy is queued. In case no ACK is received within a certain time frame, the message is re-sent.  Resending of message is attempted for a maximum of three times. If the destination is not available, then the message is routed to another queue for later delivery. Along with the implementation of new code, these APIs make use of the Netty open-source protocol framework. The implementation makes use of three packages of the Netty framework: HTTP and web sockets, sockets and datagram, and Large File Transfer package. 

Similar to the VoIP APIs, the Video APIs are also implemented on top of the open-source Mobicents Application Server platform and uses the Mobicents Asterisk package for video communication. The Video API enables reception and transmission of video between command and control station and mobile users.

\textit{Navigation APIs}: The navigation APIs collect location statistics of all the WiMesh clients and display these location specific statistics on a map on the server. The navigation APIs deal with locating specific users and, if needed, navigating to the required user(s) using Google maps API.  The GPS module in the mobile phone gets the precise location of the user in terms of latitude and longitude and then passes this information to the Client Navigation API which passes this information to the Server Navigation API.

\textit{SIP Server APIs}: The job of the SIP server APIs is to interface the WiMesh system with the OpenSIPS server. These APIs relay audio requests and data from the WiMesh clients to the OpenSIPS server. These APIs also relay SMS between WiMesh clients and the OpenSIPS server.

\textit{Communication APIs}: The communication APIs provide the communication interface between the WiMesh server and the mesh backhaul, OpenSIPS, and the clients. Any communication from other components of the system to theWiMesh server takes place through these APIs. These APIs also handle the socket creation, maintenance, and destruction.

\textit{Authentication APIs}: The authentication APIs provide user registration and authentication functionality.  The WiMesh system supports two levels of security: the first level is ``open access'' in which no registration or authentication is required; whereas the second level requires user registration and login to the WiMesh system. The password can be saved at the client side for automatic authentication for subsequent logins.

\begin{figure*}
   \centering
\begin{tabular}{cc}
\includegraphics[width=8cm,height=5.5cm]{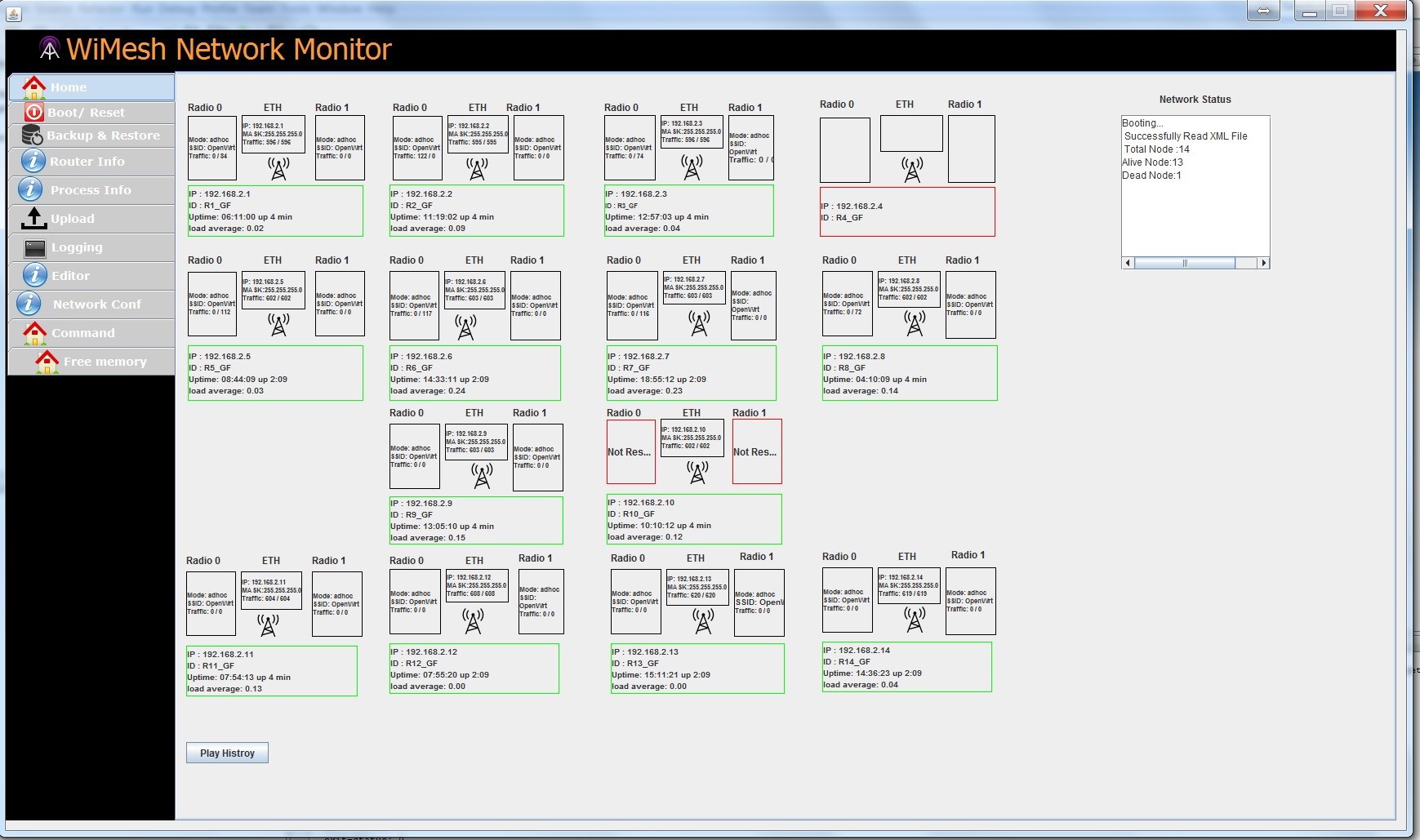} & \hspace{0.8cm}
\includegraphics[width=8cm,height=5.5cm
]{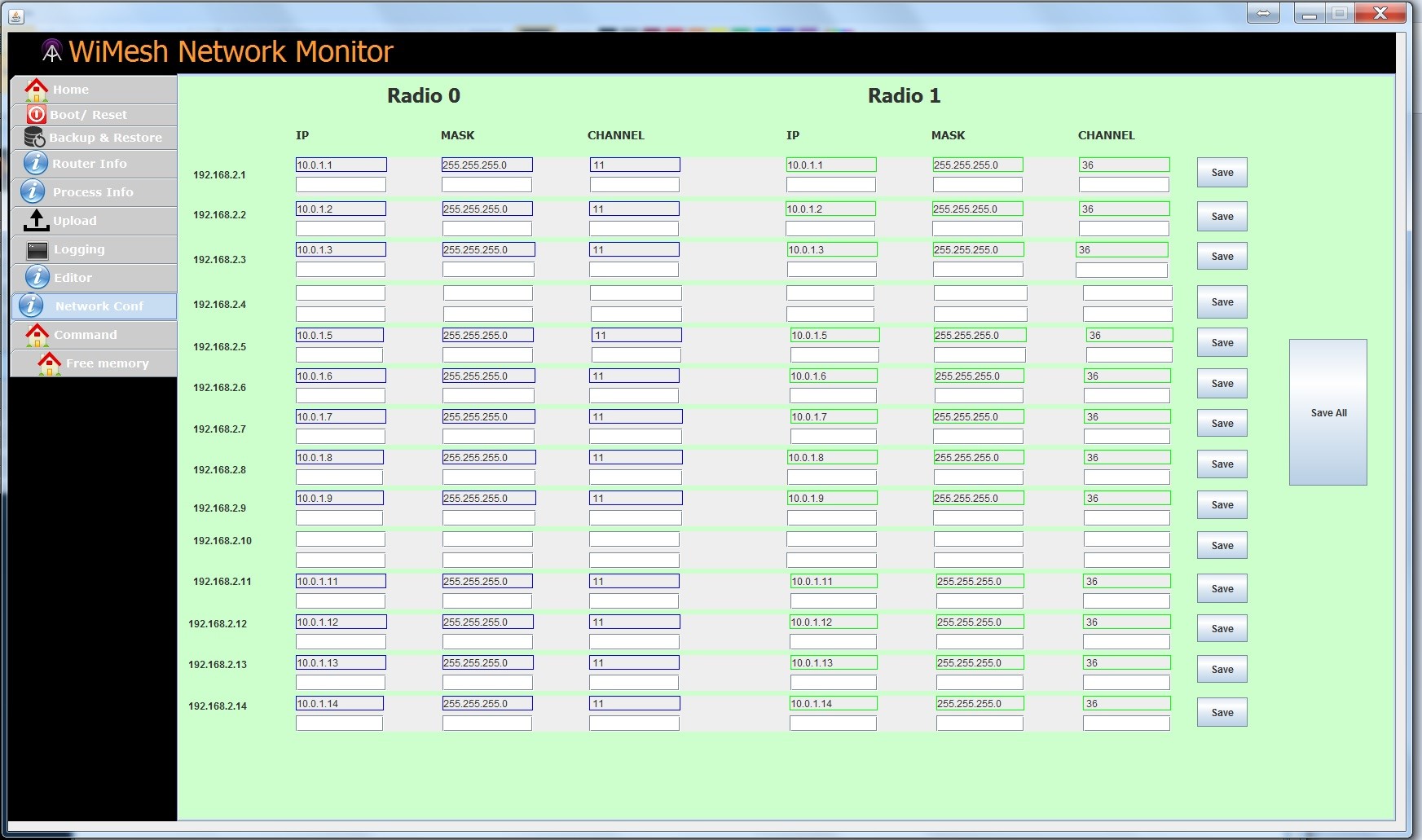}
\end{tabular}
    \caption{WiMesh Controller Homepage and Network Configuration Page}
    \label{WiMeshController} 
\end{figure*}

\vspace{2mm}
\subsubsection{WiMesh Controller}
The WiMesh controller performs mesh network management, monitoring, and logging. It provides the following salient functionalities:

\begin{itemize}
\item Listing the current nodes status (up, down, restarting); 
\item Live statistics on routes, gateway, flags, and metric;
\item Information on all the executing processes;
\item Information logs such as router ID, router IP, received packets, transmitted packets, lost packets, interface ID, interface IP, and transmission queue length;
\item Files and folders upload and download;
\item  Backup of the current state of all the network routers;
\item Network restore to a previous state;
\item Single screen remote shell access for fine-tuned control;
\item Remote editing of the files on any router; and
\item Remote modification of the radio interfaces of any router.
\end{itemize}

The WiMesh controller performs the management, monitoring, and logging of the mesh network. While there are some comprehensive and useful management tools available, the WiMesh system requirements were so diverse that no single tool met all the requirements. This motivated the development of a custom-built WiMesh controller using the Java programming language. The WiMesh controller has a user-friendly interface for monitoring the mesh network, collecting statistics, and logging network traffic. The tool uses JSch (Java Secure Channel), which is a Java implementation of the SSH2, for opening sockets to the mesh nodes. The JSch facilitates secure remote login, secure file transfer, and secure TCP/IP and X11 forwarding. The home page and the network configuration page of the WiMesh Controller can be seen in Figure \ref{WiMeshController}. 

\vspace{2mm}
\subsubsection{OpenSIPS VoIP Server}
The OpenSIPS VoIP sever is an open-source, multi-functional, multi-purpose signaling SIP proxy server \cite{opensips}. The application server interfaces with the OpenSIPS VoIP server to relay voice, text, files, and video among WiMesh clients and the OpenSIPS sever.

\vspace{2mm}
\subsubsection{The Protocol Stack}
The protocol stack is based on the Asynchronous Event-Driven Network Application Framework. The WiMesh server requires multi-threading approach to handle multiple client connections simultaneously. Research shows that the traditional multi-threading approaches perform poorly and usually incur significant overhead \cite{welsh2001seda} especially for a performance driven system.  Hence, the WiMesh system uses a more efficient approach, Staged Event-Driven Architecture (SEDA) \cite{welsh2001seda}, that avoids the high overhead associated with thread-based concurrency models and decouples event and thread scheduling from application logic. By performing admission control on each event queue, the service can be well conditioned to load, preventing resources from being overcommitted when demand exceeds service capacity.  

\subsection{WiMesh Client}
\begin{figure}[b]
\centering
\includegraphics[width=8cm, height=4cm]{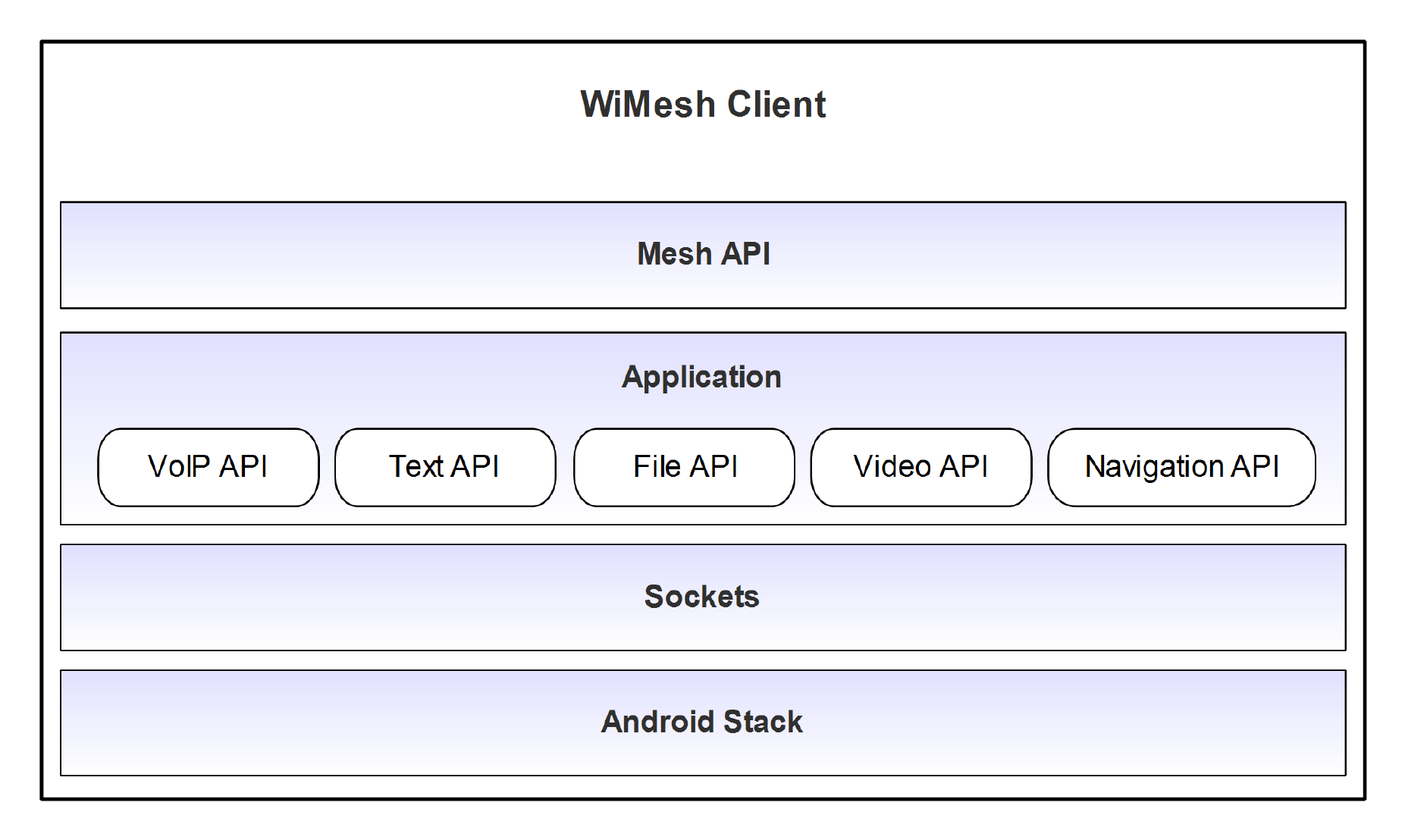}
\captionsetup{singlelinecheck=true} 
\caption{\label{ClientArch} WiMesh Client Architecture}
\end{figure}

The WiMesh client module is installed on mobile devices carried by the users and enables multimedia communication with other clients and the WiMesh server.  The client module consists of the Mesh API, the client application, socket, and the Android stack, as shown in Figure \ref{ClientArch}. 

\textit{Mesh APIs}: The Mesh API interfaces with the mesh network and is used to transmit and receive client data and control packets to and from the mesh network. The Mesh API creates message headers for packets containing client information and the services required such as voice, video, or text. The client Mesh APIs are implemented to ensure the periodic sending of local information to the server including the local IP address and the GPS co-ordinates of the clients.

The client application module includes VoIP APIs, Text APIs, File APIs, Video APIs and the Navigation APIs. 

\textit{VoIP APIs}: The VoIP APIs provide audio communication and are implemented to allow mobile users to make voice calls between each other as well with the command and control station. The VoIP APIs use the Android SIP package.

\textit{Video APIs}: The Video APIs provide video communication and are implemented to enable the transmission of captured or live video feeds to the command and control station as well as to the other users. The video APIs use the services of the Android media package for video communication.

\textit{Text and File APIs}: The Text and the File APIs provide text communication and file upload/download capability between the mobile users as well as with the command and control station.  Both APIs use UDP and TCP sockets along with Android HTTP package for communication. UDP socket is used for efficiency. However, to ensure reliability in the case a UDP packet is lost, code is implemented so that the destination sends an ACK for a text message received. In the case, the ACK is not received within a timer period, the packet is considered lost and the application layer asks the UDP to re-send the packet again. Reliability is also included for file transfer over the unreliable UDP. Since the packet will be relayed by the server to the destined mobile user, a custom packet is used in which the IP address of the eventual mobile user is included along with the IP of the server to which the packet will be forwarded as the next step.

\textit{Navigation APIs}: The Navigation APIs periodically send GPS coordinates of the client to the server. The Navigation APIs are built on top of the Google Maps APIs.

The socket and the Android stack are already implemented modules and the aforementioned APIs will use their services through well-defined interfaces. The socket provides the interface for networking, enabling the client to communicate with other clients and the server over the mesh.

\subsection{WiMesh Backhaul Network}
The core of the WiMesh system is formed by the wireless backhaul network which creates the wireless multihop backbone, enabling mobile users to communicate with each other spread over a region. There were several aspects related to the design of the wireless mesh backhaul network.  The next subsections describe the design choices made and the motivation for those choices.

\vspace{2mm}
\subsubsection{Mesh Nodes and Associated Hardware}
The project uses two different types of mesh nodes: PicoMesh node and NanoMesh node. The mesh hardware selected for the project was the Ubiquiti PicoStation M2HP and Nano-Station M2 models \cite{ubiquiti}. The PicoStation is a 600mW, 2.4 GHz, high power outdoor access point, equipped with an Atheros MIPS 24KC, 400 MHz processor, 32MB SDRAM, 8 MB Flash, one 10/100 Ethernet port, and a 6 dBi outdoor omni-antenna. Weighing a mere 100 g, with dimensions of 136x20x39 mm, and the ability to withstand extreme temperatures (-20 \degree C to 70 \degree C), the PicoStation is portable, rugged, and provides omni coverage of up to 1 km in open areas while consuming only 6 Watts. This specification makes the PicoStation a good choice for use in meshing or as an access point to which mobile nodes can connect. The NanoStation is a more powerful device comprising of a 200mW 2x2 MIMO radio with an integrated dual polarity 13 dBi sector antenna, a horizontal beamwidth of 90, and providing outdoor coverage of up to 10 Km.

Every node comprises of an extensible stand (up to 25 feet), a 12 Volt, 7 AH dry battery, a small TP-Link (model TL-SF1005D) Switch, and modified Power over Ethernet (PoE) units (to enable direct connection to 12 V battery instead of 220 V DC supply) or 12 Volt PoEInjector units to power the mesh node. In addition, portable custom-made mesh-boxes are used at each node to house the various hardware components and for protection from weather conditions. The server mesh node additionally hosts the WiMesh Server in addition to the standard hardware.

\begin{figure}[t]
\centering
\includegraphics[width=6.5cm, height=6.5cm]{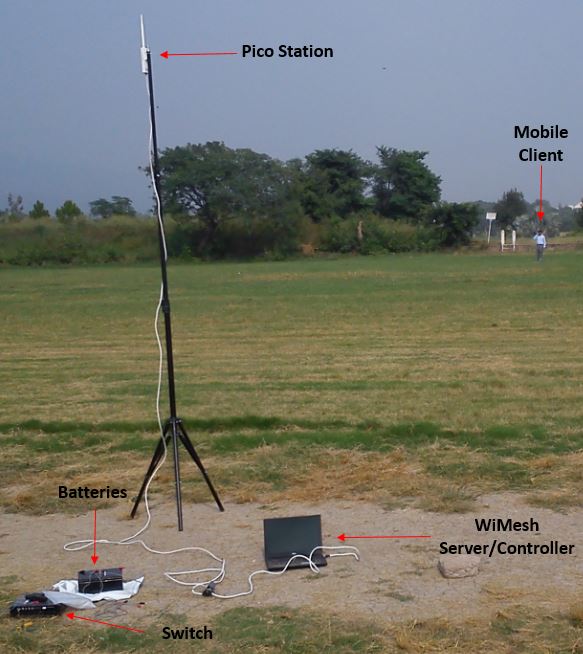}
\captionsetup{singlelinecheck=true} 
\caption{A typical PicoMesh node}
\label{PicoMeshNode}
\end{figure}

The PicoMesh node comprises of two PicoStation devices connected through the switch. One PicoStation at each node operates in the access point mode and allows mobile devices to connect. The second PicoStation operates in the ad-hoc mode, runs the mesh OLSR protocol, and forms the wireless multihop backbone. The two devices are tuned to non-overlapping channels to avoid interference between the mesh backbone and the access network. The PicoStation has a relatively limited range but provides omnidirectional coverage. Hence, the PicoMesh node is used when a higher degree of meshing is required in a smaller area with problems of line-of-sight on the terrain, requiring multiple alternate routes to bypass obstacles such as dense foliage and small mountains.

The NanoMesh node comprises of two NanoStation devices and a single PicoStation device connected through the switch. Since the NanoStations provide sectorized wireless coverage, two devices are required at a minimum to enable the wireless reach to other nodes. Another noteworthy difference from the PicoMesh is that NanoMesh requires some alignment due to the sector coverage whereas the PicoMesh provides omni-coverage and does not require the deployment positions to be constrained if the mesh node is within the coverage of at least another node. As in the case of PicoMesh node, the PicoStation device is used to provide access to mobile devices since it offers omni-coverage. The NanoMesh node is the node of choice for covering large regions with as few nodes as possible. This is the typical usage pattern for rural areas or for disaster situations, in which a large area needs to be covered with a clear line-of-sight using a few nodes. Together, the two NanoStation devices can connect mesh nodes separated as far as 20 km. Figure \ref{PicoMeshNode} shows a typical PicoMesh  node with batteries, switch, associated cabling, the WiMesh server/controller, and a mobile client. The server and the controller housed by the laptop was later replaced by Intel NUC device. The mesh-box (not shown in figure) is waterproof, rugged, and enables easy portability and deployment of the WiMesh System.

\begin{figure}[t]
\centering
\includegraphics[width=7cm, height=7cm]{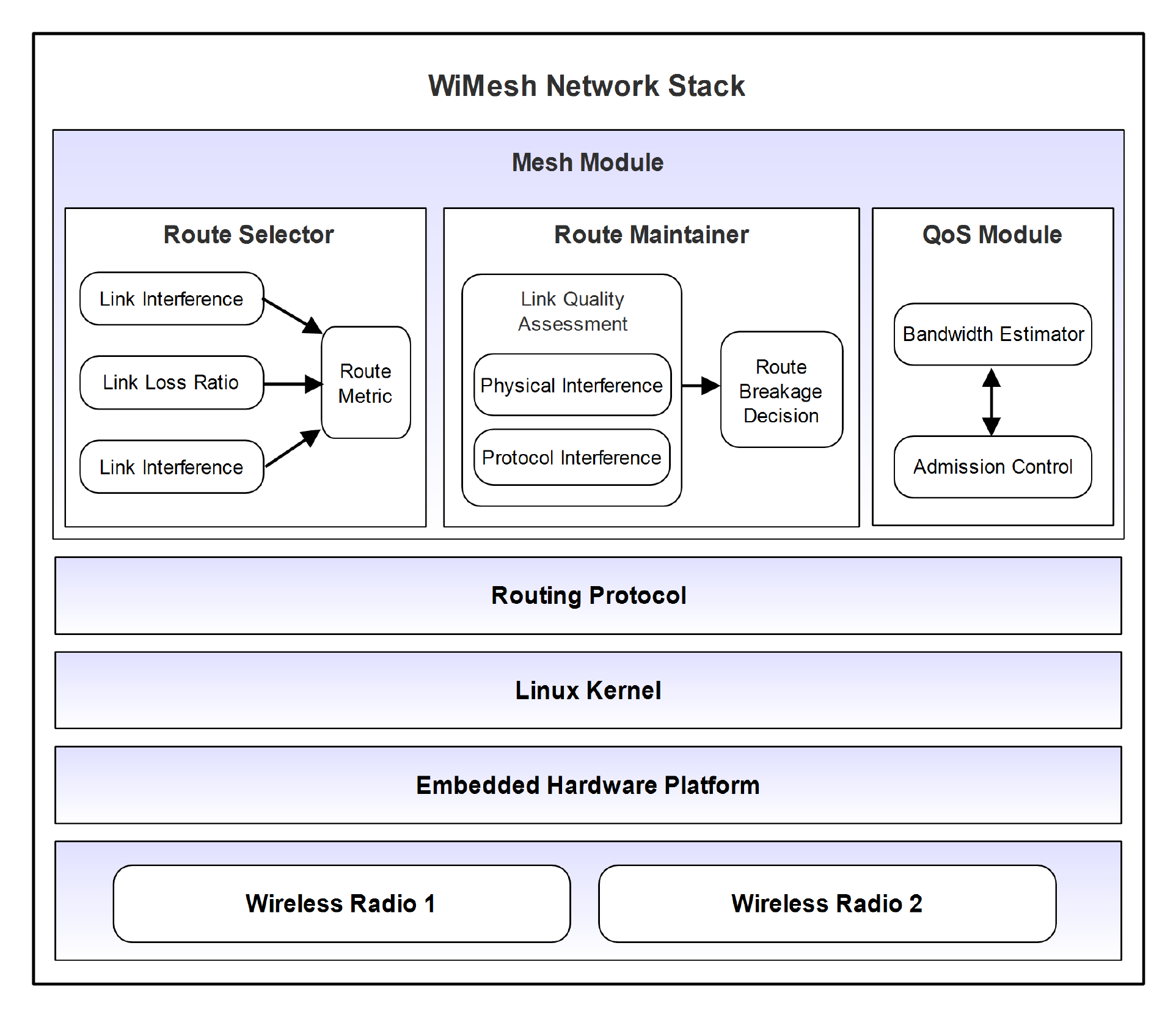}
\captionsetup{singlelinecheck=true} 
\caption{WiMesh Network Stack}
\label{MeshNetworkStack}
\end{figure}
\begin{figure*}[t]
   \centering
\begin{tabular}{cc}
\includegraphics[width=8.5cm,height=6cm]{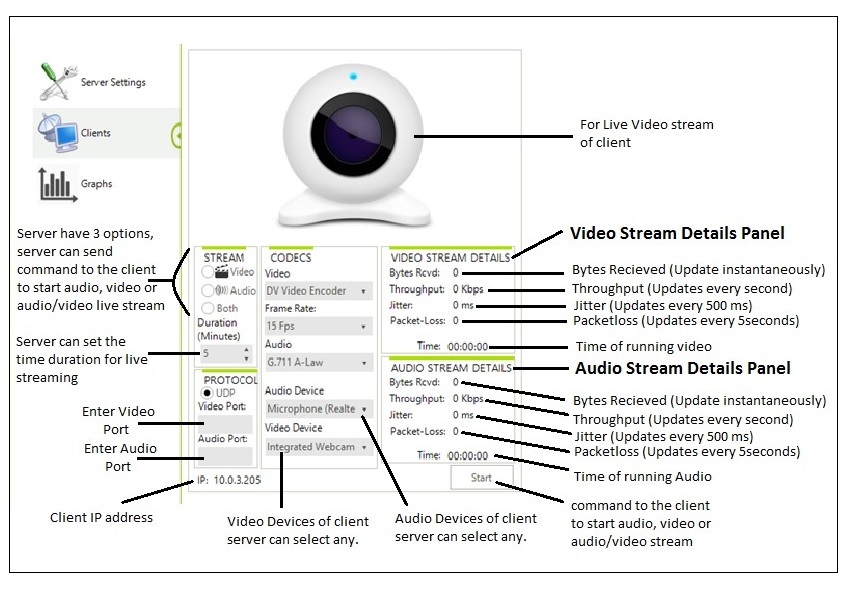} &
\includegraphics[width=8.8cm,height=6.2cm]{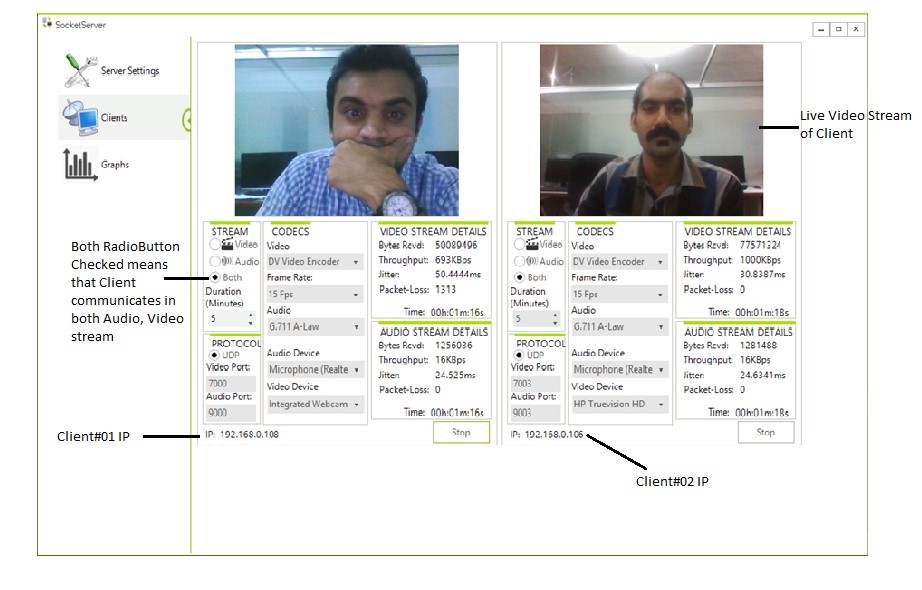}
\end{tabular}
\caption{Sample Multimedia Performance Evaluation Tool Session}
\label{MultiMediaPerfTool} 
\vspace{-0.5cm}
\end{figure*} 

\vspace{2mm}
\subsubsection{Mesh Performance Metric and Routing Protocol}
Figure \ref{MeshNetworkStack} shows the WiMesh backhaul network stack which consists of a Mesh Module on top of a Linux kernel running on an embedded hardware platform. The performance of the mesh module determines, to a large extent, the overall performance of the system. The primary job of the Mesh Module is to interconnect the clients and the server over a wireless multihop infrastructure. There are three main sub-components of the Mesh Module: Route Selector, Route Maintainer, and the QoS Module. While there are several areas where performance can be improved for a mesh network, two areas are particularly important from the perspective of providing QoS for multimedia traffic: route selection and route maintenance. 

\textit{Resilient Route Selector}: The route selection portion deals with selecting stable, high performing end-to-end communication routes in the mesh network between the source and destination pair to provide an acceptable level of quality of service (QoS) for multimedia flows. Route selection has been shown \cite{de2005high} to bring substantial performance improvements, particularly in the achievable end-to-end throughput. The traditional hop count metric is not a good approach \cite{de2003performance} and a realistic metric needs to consider multiple factors pertaining link quality. In our previous contribution, the ELP metric \cite{ashraf2013route} was proposed which addresses several limitations of existing routing metrics including asymmetry of the links, radio interference, logical interference, as well as inter-flow and intra-flow interference. ELP solves the problems of asymmetry and inaccurate probe-approximation by introducing a corrective constant in the calculation to bias the estimation in favor of link loss ratios in the forward direction (data packets). 

ELP explicitly incorporates the Channel Contention Interference (CCI) arising from the CSMA-CA based MAC by using the promiscuous listening mode \cite{yang2005contention, chen2005qos, kim2006accurate} of IEEE 802.11. A node can observe the channel states using the NIC card. ELP also takes into consideration the link capacities as higher capacity links can transmit data at a higher rate and, therefore, occupy the medium for a shorter period. Low capacity links, however, take longer time and will create interference for nodes in vicinity and even for far away nodes that are outside carrier sensing range. In ELP, links with higher bandwidth (radio transmission rate) are given a lower link cost. Further, a few existing routing metrics cater to the problems which can be caused by frequently changing routing metrics which cause excessive routing overhead and frequent route shifting (or route oscillations). This is avoided in ELP in addition to offering quick recovery from failed nodes in ELP to offer a more resilient communication solution. Formally, the ELP metric for a link \textit{e} is defined as:
\begin{center} 
\textit{ELP(e) = Link Loss Ratio(e) $\times$ Link Interference(e) \\
$\times$ Link Capacity(e)}
\end{center}

The ELP metric for a path \textit{p} is defined as:
\vspace{-0.5cm}
\begin{center} 
$$\textit{ELP(p)} = \sum_{link \thinspace e \in p} \textit{ELP(e)}$$
\end{center}

Overall, ELP enables the selection of high performing end-to-end routes across the mesh network, enabling better performance for multimedia flows. More details about the proposed routing metric ELP can be found in \cite{ashraf2013route}.

\textit{Route Maintainer}: The route maintenance component deals with maintaining end-to-end routes for the duration of the flow. In CSMA-CA based wireless multihop networks, such as IEEE 802.11, routes are fragile and often break down because of transitory interference problems on any of the intermediate wireless links \cite{ashraf2011route}.  These problems are particularly severe when several flows traverse the network, creating a high degree of interference. Route maintenance aims to provide stability to the end-to-end route despite the link instabilities of wireless links. Efficient route maintenance can provide substantial performance improvement in mesh networks in terms of reducing route breakages. In IEEE 802.11 based networks, the standard mandates a retry limit of seven \cite{lan2003part}. If there are eight consecutive failed transmission attempts (one try and seven retries) to the next-hop node, then the link layer sends a failure notification to the network layer. Protocols, such as AODV, systematically translate a link breakage into a route breakage and the intermediate node notifies the source node to find new routes. The Route Maintainer mitigates this problem and considers the long-term link performance \cite{ashraf2011route} to offer a resilient solution.

\textit{QoS Module}: The main job of the QoS module is to keep an estimate of the available bandwidth in the wireless mesh network and to perform admission control for incoming flows based on the bandwidth availability on the wireless multihop path. This bandwidth management is necessary because although the system is designed to accommodate the maximum number of flows possible, too many flows can overwhelm the wireless mesh networks and cause performance degradation for existing flows. Bandwidth estimation is primarily based on a hybrid approach involving using a model as well as actual channel measurements using promiscuous listening at the data link layer to be as accurate as possible.

\subsection{Additional WiMesh System Components}
In addition to the main components, the WiMesh system consists of some auxiliary components.  One such component is the performance evaluation tool for the multimedia traffic of the WiMesh network. Performance evaluation tools are important in designing, evaluating, and analyzing the performance of multimedia traffic of mesh-based wireless network due to the high bandwidth requirements of such traffic.

This performance tool is installed on end stations connected to the mesh network. The end stations are connected to cameras or microphones.  This tool is used for streaming live multimedia traffic (audio and video) from connected clients to the server and analyzing statistical data of these live streams in a wireless mesh network.  The server aggregates data from all video/audio flows.  The statistical data consist of throughput, delay, jitter, packet loss, video quality, and audio quality and can be visually viewed in terms of graphs.  The server is aware of the connected clients and can separately send remote commands to each connected client to (1) select video device of the client; (2) select video codec of the feed; (3) select frames per second of the video feed; (4) select audio device; (5) select audio codec; and (6) start or stop the live feed.

The tool provides session management and data logging capabilities for scheduling experiments and storing the traffic metrics during the actual transfer of the data. The results obtained are used in the final evaluation of mesh routing protocols, metrics, and other communication related mechanisms.  The tool uses UDP protocol at the transport layer. Figure \ref{MultiMediaPerfTool} shows two snapshots of the performance tool session.

\section{WiMesh System Deployment Experience}
Several tests were carried out on both indoor and outdoor setup to evaluate the functionality and performance of the WiMesh system. More than a dozen of field trials were conducted at different times and places. Moreover, due to the diverse equipment involved, several outdoor visits were required to properly setup, power the devices, and to ensure proper interconnections between components.

\subsection{Indoor Testbed}
This section describes the experience gained during the deployment of the indoor testbed for the WiMesh system as well as the results of the experimental performance evaluation of the WiMesh system.

\vspace{2mm}
\subsubsection{Testbed Environment}
Before designing a testbed, it is essential to physically survey the site to determine its suitability to host the proposed testbed. There are several factors to consider:

\begin{itemize}
\item \textit{Offer easy physical access}: The actual act of deployment involves several physical activities including installing wires, putting the mesh node in place, and securing the enclosures among others. Moreover, nodes in any testbed are known to be error-prone and it is sometimes required to remove a node or change its connections;
\item \textit{Offer easy wired access}: Despite being a wireless network, wired connection is eventually needed from a central switch to each node to facilitate controlling and manipulating the nodes remotely. This is an important constraint because installing wires require ducting, drilling, and administrative permissions. Moreover, the costs become higher as the length of deployed wires become longer. Thus, locations of nodes were selected for ease of wired installation and proximity to existing ducts where possible;
\item \textit{Offer protection from weather conditions}: Weather protection was important as the proposed testbed was not entirely in rooms, but rather in corridors. Therefore, it was  necessary to select a location where devices are protected from rain and direct sunlight during summers;

\item \textit{Offer relative protection against theft}: It was important to select locations where normal students or visitors cannot easily access the device physically. This does not mean that nodes must be in a very difficult to reach place, however, the selected location should require the help of ladders and other tools to access the device. Some excellent locations had to be left out because the node would become easily approachable by visitors or students.
\end{itemize}

Based on the above factors, most of the nodes were deployed on the three-storey A-Block of the Computer Science Department of Air University, Islamabad, Pakistan.

\vspace{2mm}
\subsubsection{Testbed Design}
\begin{figure}
   \centering
\begin{tabular}{c}
\includegraphics[width=7cm,height=4cm]{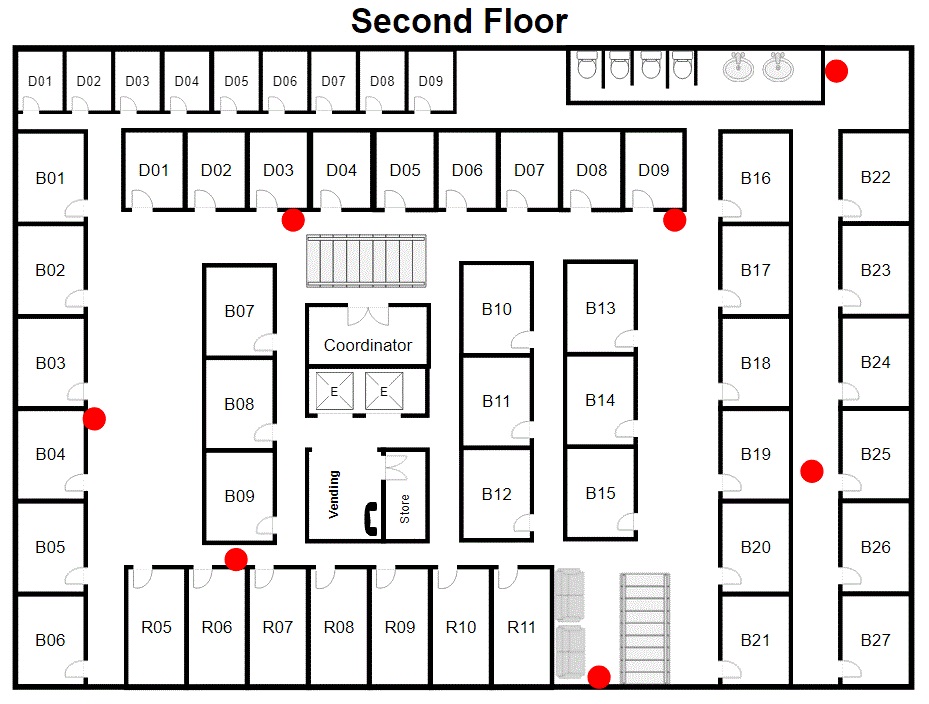}\\
\includegraphics[width=7cm,height=4cm]{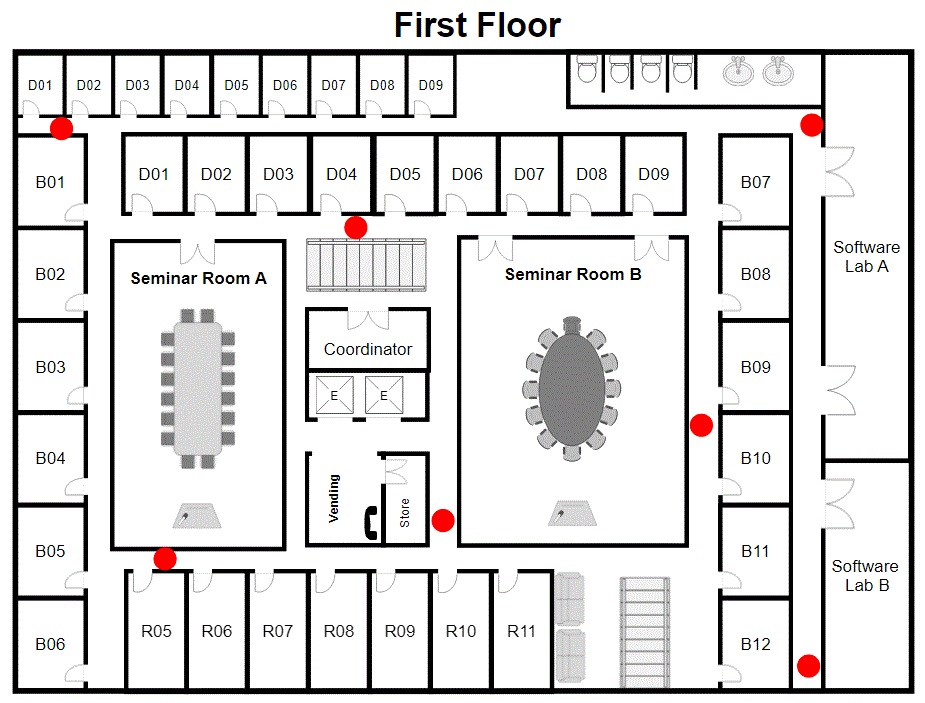}\\
\includegraphics[width=7cm,height=4cm]{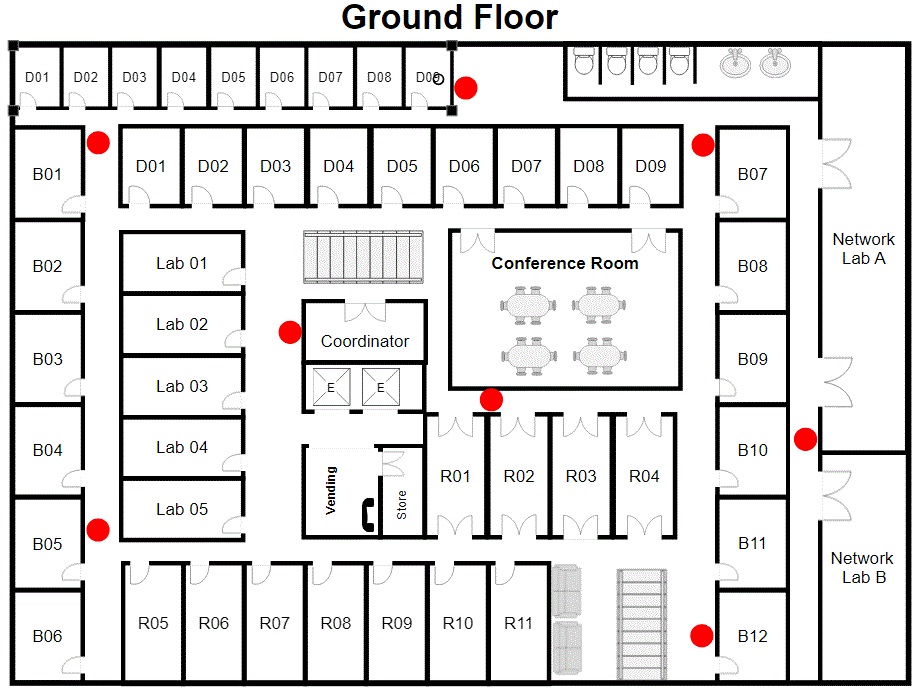}
\end{tabular}
    \caption{Routers deployment in the A-Block of Air University}
\label{InstallmentPositions} 
\vspace{-0.5cm}
\end{figure}

A 43 node testbed was deployed, which is a decent size for indoor wireless experimentation. The nodes were distributed across different departments of the Air University, however, the larger concentration of nodes (22 out of 43 nodes) was in A-Block, as shown in Figure \ref{InstallmentPositions}. TP-Link WDR4300 routers, running OpenWRT embedded Linux firmware, based on IEEE 802.11n (MIMO) were used with data rates of 300 Mbps. The Power-Over-Ethernet (PoE) technology was used to power each of the routers since installing power sockets was impractical and dangerous. 

\vspace{2mm}
\subsubsection{Deployment of the Testbed}
Deploying the indoor testbed was a long and grueling process that involved several aspects including purchase of diverse equipment, supervision of on-site labor and technicians, and administrative logistics. Table \ref{IndoorTestbedSpecs} shows the testbed specifications.

\begin{table}[H]
\caption{Indoor Testbed Specification}
\centering
\label{IndoorTestbedSpecs}
\begin{tabular}{|l|l|}
\hline 
\cellcolor{black!25}\textbf{Parameters} & \cellcolor{black!25}\textbf{Description/Value}\tabularnewline \hline
\begin{tabular}[c]{@{}l@{}}Number\\   of Nodes\end{tabular} & 22 \\ \hline
\begin{tabular}[c]{@{}l@{}}Covered\\   Floors/Area\end{tabular} & \begin{tabular}[c]{@{}l@{}}3  Floors;  Covered Area = 16000 sq. ft\end{tabular} \\ \hline
\begin{tabular}[c]{@{}l@{}}Node\\   Hardware\end{tabular} & \begin{tabular}[c]{@{}l@{}}TP-Link WDR4300, \\Dual-Radio, Dual-Band (2.4 GHz and 5 GHz), \\ routers with aggregate throughput\\of up to 750 Mbps, 2 USB ports\end{tabular} \\ \hline
\begin{tabular}[c]{@{}l@{}}Frequency\end{tabular} & 2.4 GHz and 5 GHz \\ \hline
\begin{tabular}[c]{@{}l@{}}Power Source \end{tabular} & Power Over Ethernet \\ \hline
\begin{tabular}[c]{@{}l@{}}Operating System \end{tabular} & OpenWRT\\ \hline
\end{tabular}
\end{table}

Since the design of the complete system was well thought-out, the integration of the WiMesh system with the testbed was straightforward. A laptop or any other device was attached which was configured with the WiMesh server and the associated services and software. The attached device was used to monitor the multihop capability of the mesh testbed to see how it performed. There were some issues regarding the inter-operability of the mesh testbed technologies and the WiMesh server, however, it was handled appropriately.

\vspace{2mm}
\subsubsection{Lessons Learned}
\label{lessonsLearned__indoor}
Despite comprehensive planning, closely overseeing the project, and a ``hands-on'' approach, a few problems arose:

\begin{itemize}
\item It was observed that using a monitoring software like the mesh controller was extremely useful since sometimes wireless interfaces on routers malfunctioned or had configuration settings problems, leading to routing and performance problems. 

\item It is important to have error-checking scripts which periodically checked active routers for a possibility of ``bricking,'' i.e., sometimes due to erroneous settings, the ports inside a router became internally connected in a loop and made access impossible. This was especially true when a reboot/shutdown remote call was given for routers. To mitigate this, scripts were installed which queried the routers periodically and removed those specific bridges. 

\item Wiring was a surprisingly difficult issue since using a star topology would have resulted in several kilometers of wires. Therefore, an efficient cabling plan was used. A related problem was that the Ethernet wire has a physical limit of carrying PoE signals for up to 100 meters. To solve this problem, switches were installed at required locations that acted as repeaters to regenerate the signals, especially between the buildings. 
\end{itemize} 

\vspace{2mm}
\subsubsection{Testbed Results}
\begin{figure}
\begin{tabular}{c}

\subfigcapskip 2pt \subfigure[Packet Delivery Ratio]{
  \includegraphics*[width=7.1cm, height=4.6cm]{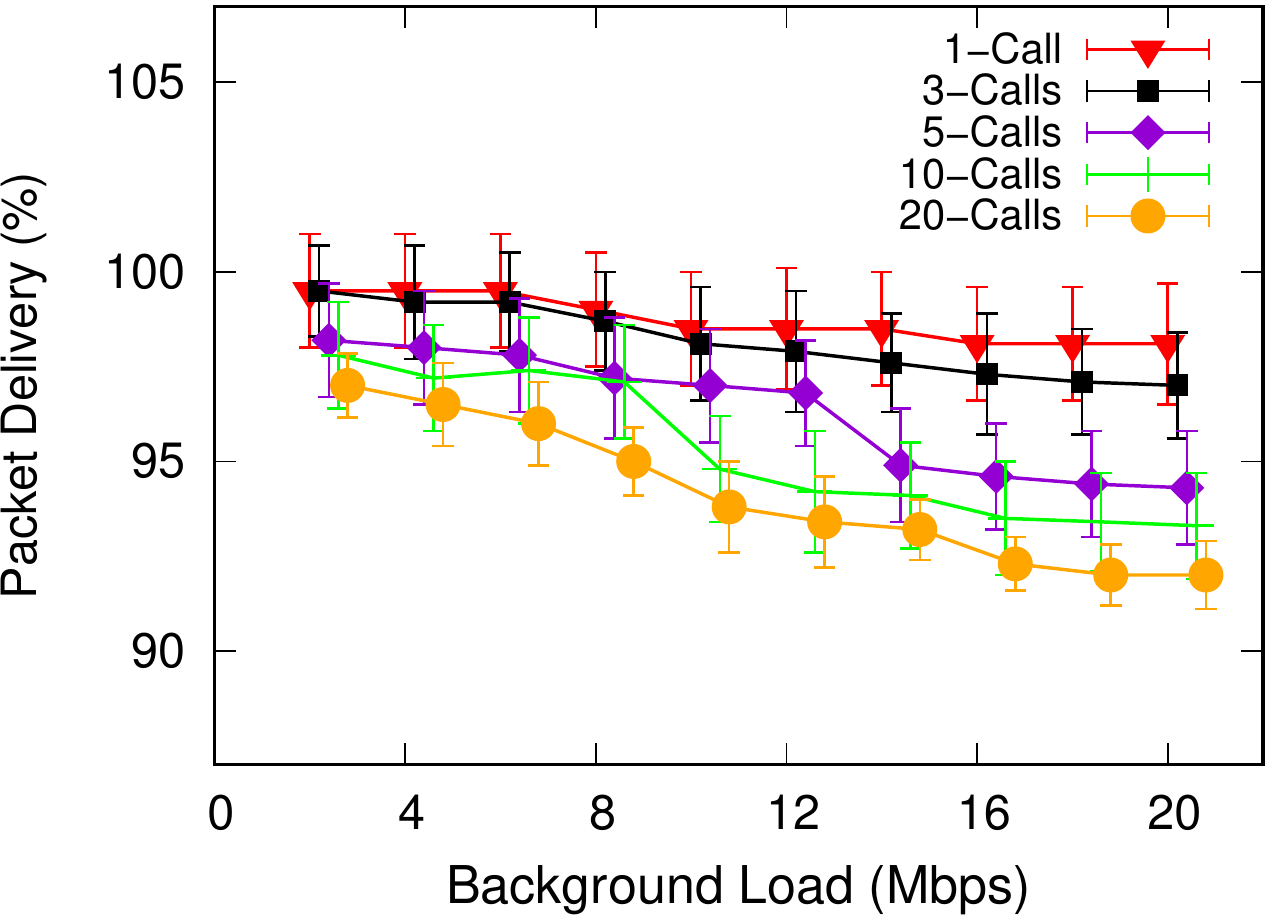}
   \label{Unevena}
   }
   \\
   \subfigcapskip 2pt \subfigure[Delay]{
  \includegraphics*[width=7.1cm, height=4.6cm]{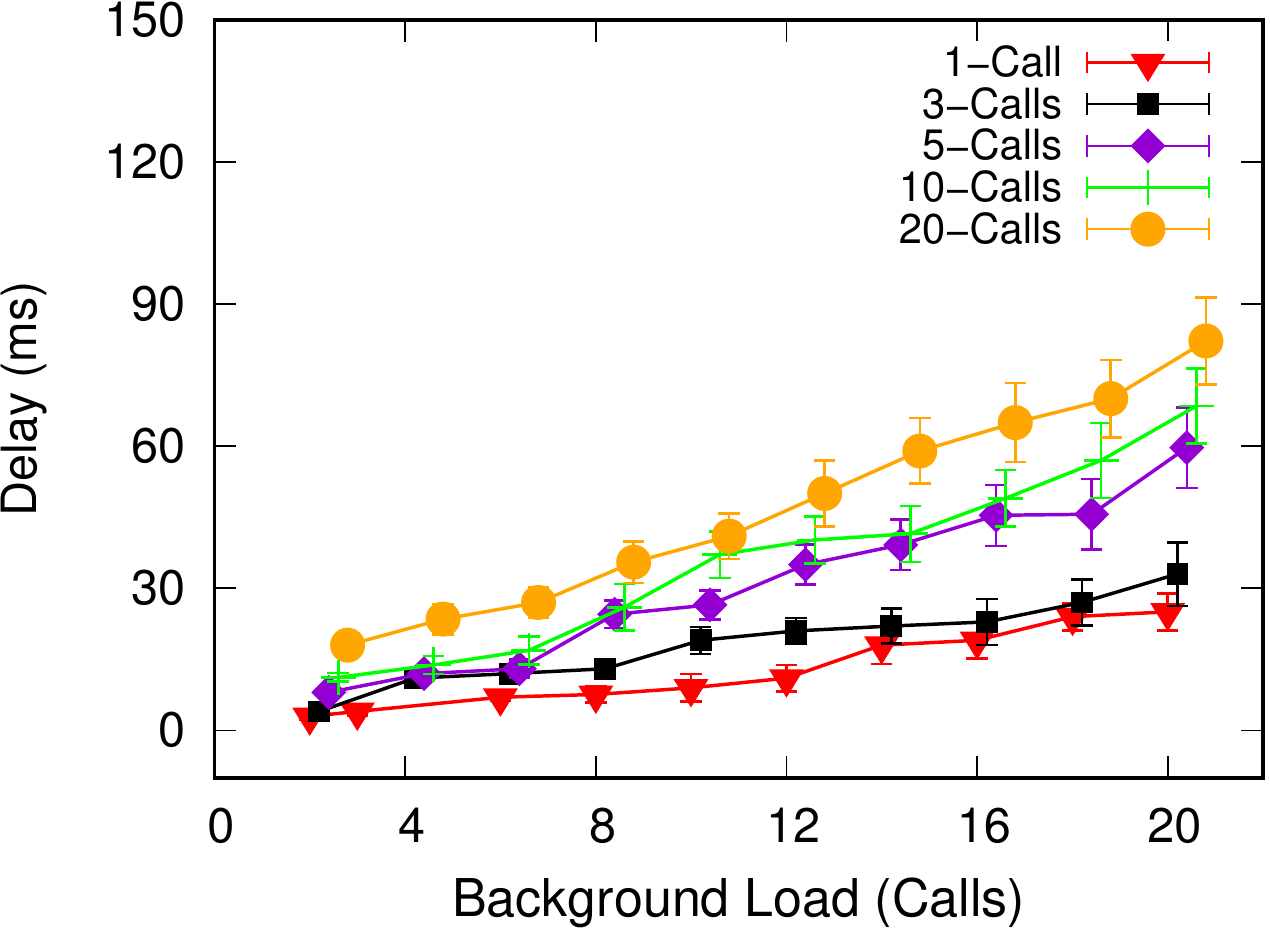}
   \label{Unevena}
   }
\end{tabular}
    \vspace{-0.1cm}
\captionsetup{singlelinecheck=true} \vspace{-0.25cm} \caption{\label{IndoorMetrics}
     \footnotesize{Indoor Testbed Performance Metrics}}
\vspace{-0.5cm}
\end{figure}

Figure \ref{IndoorMetrics} shows the performance evaluation results of the WiMesh system on the indoor testbed. The graphs have been plotted for 95\% confidence intervals to show the variation of results across different scenarios. Figure \ref{IndoorMetrics}(a) shows the Packet Delivery Ratio (PDR) for 1-20 calls while varying the background load from 2-20 simultaneous calls. As seen in the figure, the performance of the WiMesh system stays acceptable, i.e., $PDR > 90\%$ even when a reasonable load of 40 concurrent calls was placed. There is a dense mesh deployment in the experiment leading to a significant number of alternate wireless links due to both, the large number of nodes, and the availability of multiple radios and channels. The high density and multi-channel availability result in high performance of the WiMesh system in terms of PDR. Similarly, as shown in Figure \ref{IndoorMetrics}(b), the end-to-end delay is also within acceptable bounds, i.e., $ < 100 ms$. The delay becomes slightly bigger for higher number of calls since the traffic load increases compared to the lesser load scenario. Overall, the results show promising performance.  

\subsection{Outdoor Testbed}
This section describes the experience during the deployment of the outdoor testbed as well as performance evaluation of the WiMesh system.  This section presents the results obtained from the testbed deployed in the Fatima Jinnah park, located nearby the Air University, Islamabad, Pakistan. The park was selected as the site for most of the field trials due to geographic proximity and ease of equipment portability. However, insights obtained from field trials in other settings are also discussed.

\vspace{2mm}
\subsubsection{Testbed Environment}
For the outdoor testbed, the main concern was to ensure that performance evaluation of the system was carried out over a relatively large geographic region with varying terrain. Two nodes located on rooftops of Air University were also included in the testbed to extend the wireless range over a larger region. 

\vspace{2mm}
\subsubsection{Testbed Design}
Due to logistic restrictions, the experiments were carried out for a seven node testbed. Based on varying needs of wireless antenna height, an extensible stand (up to 14 feet) was used. However, light-weight stands were used for portability. Wireless coverage was an important issue and both PicoStations and NanoSations were used for outdoor mesh nodes depending on their utility. For mesh nodes with NanoStations, three nodes were deployed in circular pattern to cover wider area. A random topology was used with five nodes deployed randomly in the park while two nodes were deployed on Air University rooftops, as shown in Figure \ref{OutdoorFJP}. In order to ensure relatively large distance between nodes, the average distance between nodes was around 0.75 km with some wireless links spanning 2 km. Similar to the indoor testbed, the  OpenWRT embedded Linux OS was installed on the PicoStations and NanoStations. PoE was used to power the mesh nodes. This decision was even more pertinent for the outdoor testbed in order to avoid the dangers of live electricity, especially on wet terrains. Dry portable batteries were used to power outdoor mesh nodes in order to avoid battery refilling and spillage. Custom-built outdoor cabinets were used for the rooftop nodes since these installations were permanent and required protection from weather. The cabinets were installed with horizontal slits on sides for ventilation.  

\begin{figure}[t]
\label{OutdoorFJP}
\centering
\hspace{0.1cm}\includegraphics[width=8cm,height=5.5cm]{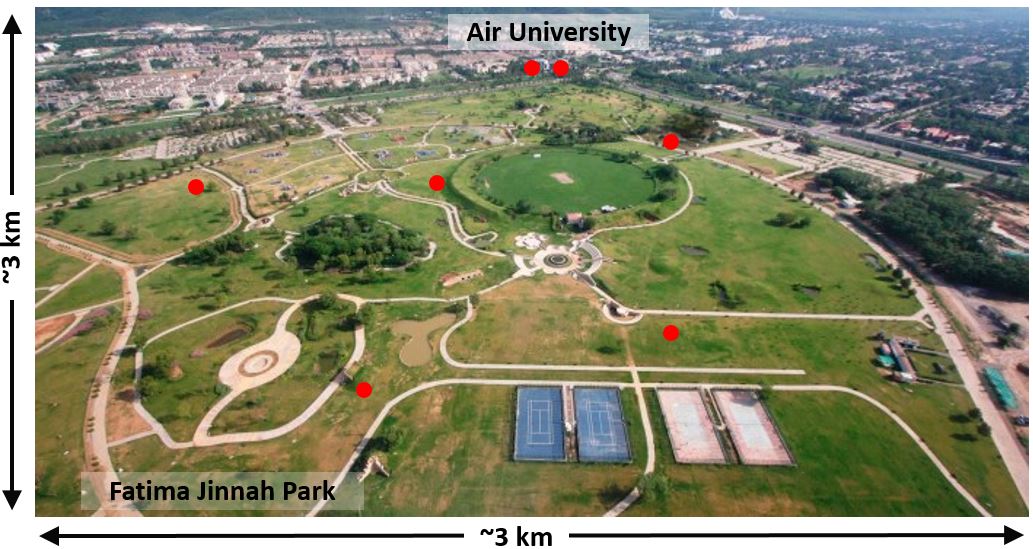}
\caption{\label{OutdoorFJP}Output Testbed Setup}
\vspace{-0.3cm}
\end{figure}

\begin{figure*}
\begin{center}
\begin{tabular}{c c}

\subfigcapskip 2pt \subfigure[Packet Delivery Ratio]{
  \includegraphics*[width=7.6cm, height=4.8cm]{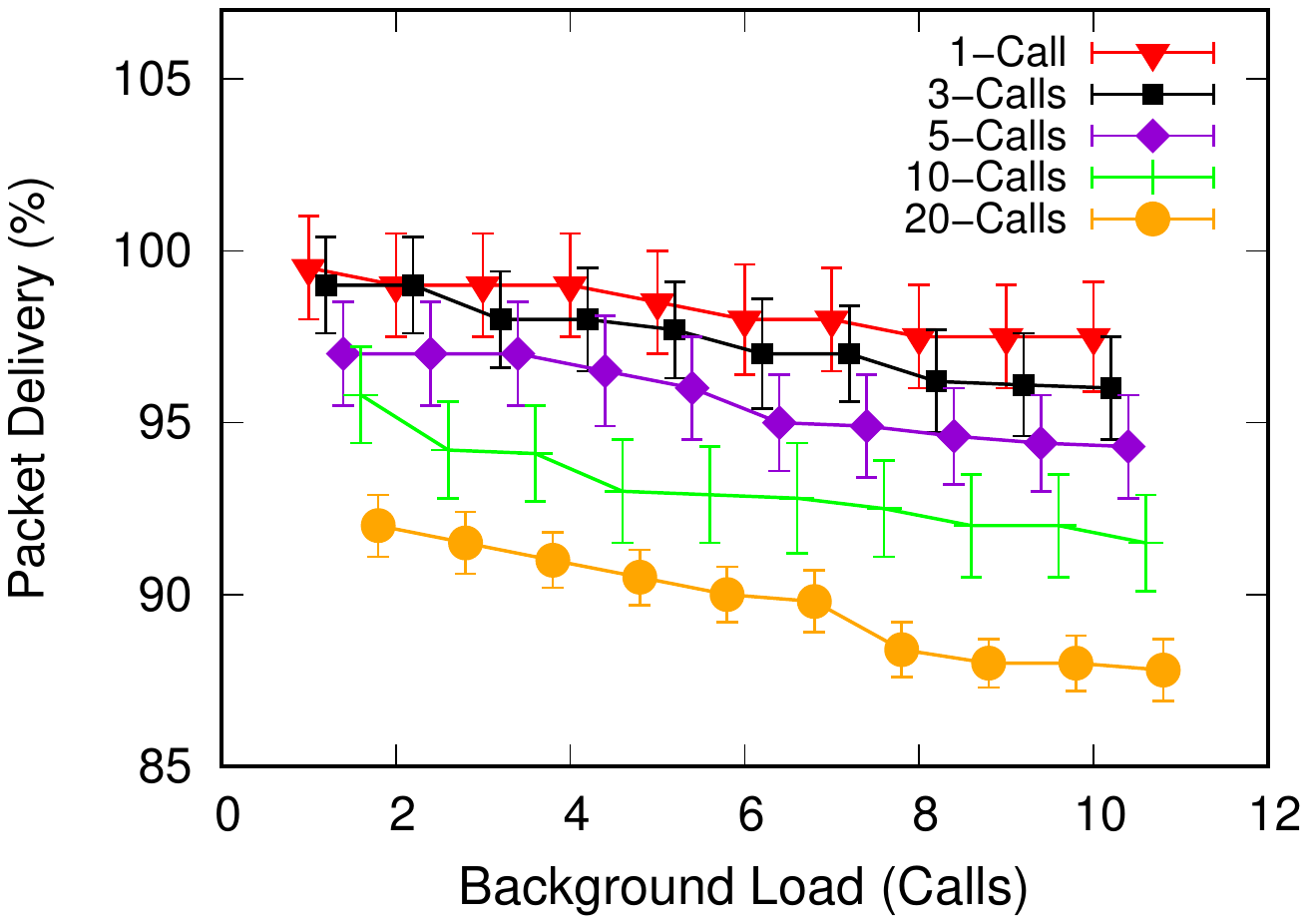}
   \label{Unevena}
   }
   \hspace{0.2cm}
\subfigure[Packet Loss Ratio]{
  \includegraphics*[width=7.6cm, height=4.8cm]{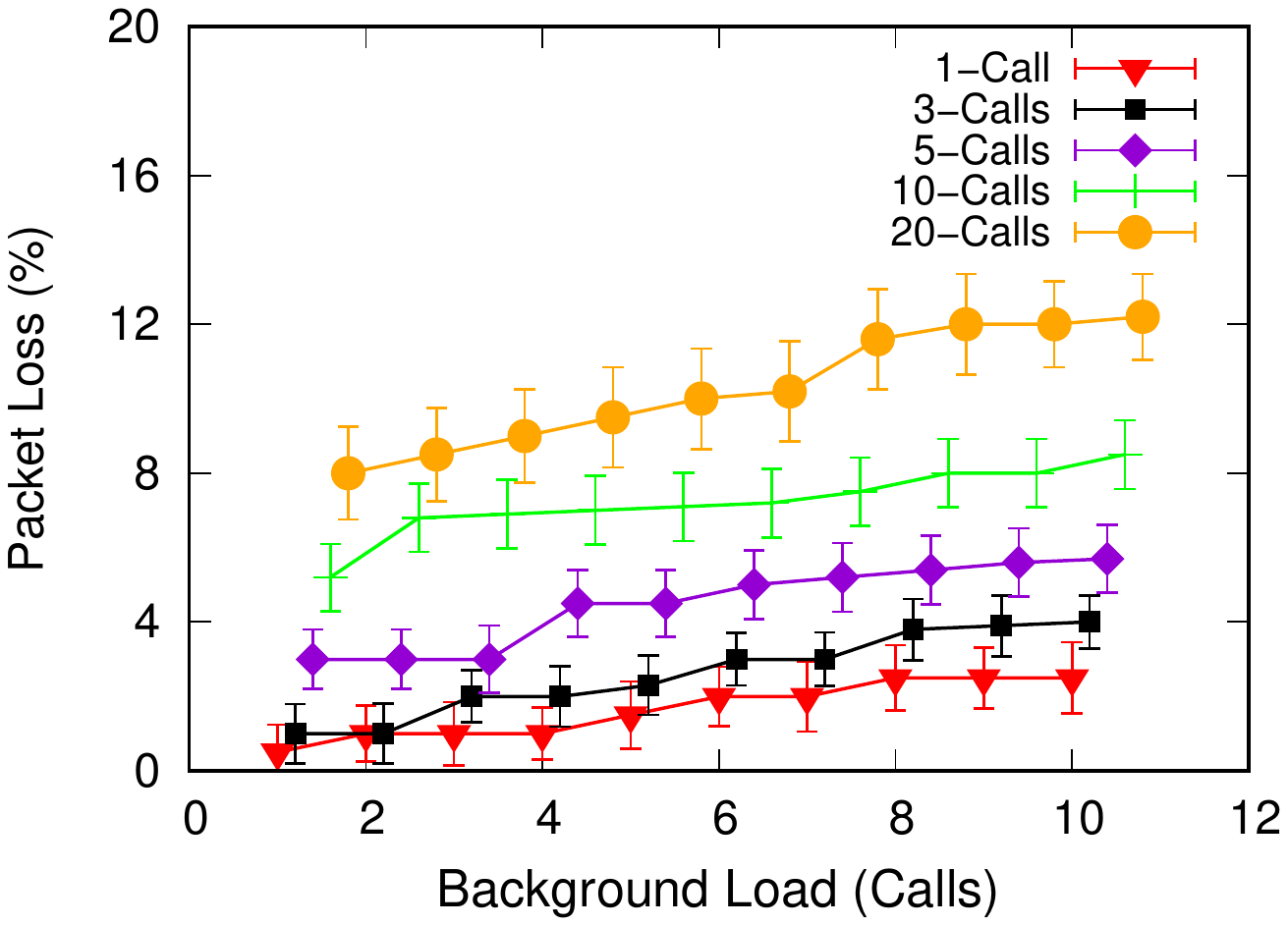}
   \label{Unevenb}
   }\\
   \subfigcapskip 2pt \subfigure[Delay]{
  \includegraphics*[width=7.6cm, height=4.8cm]{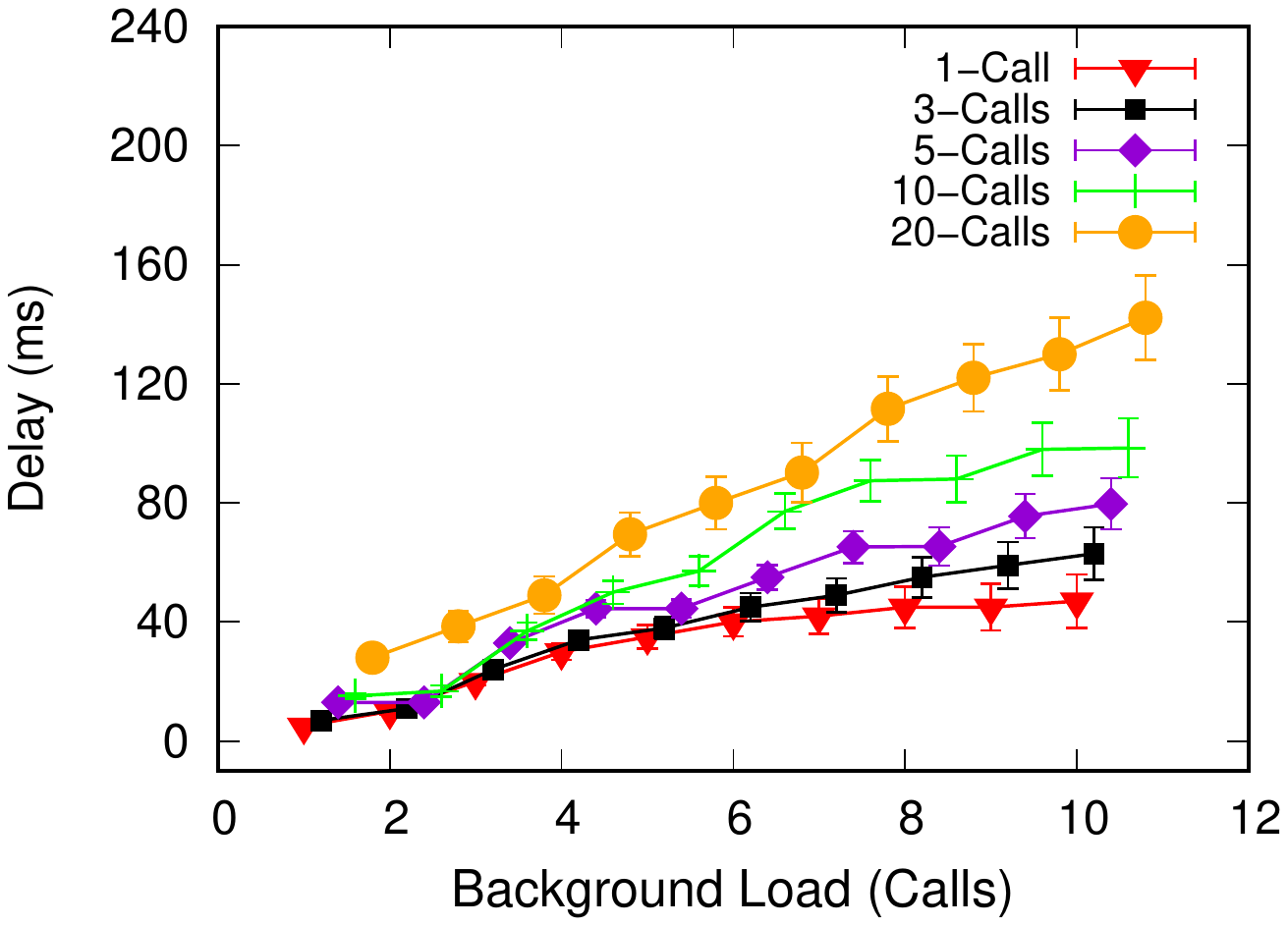}
   \label{Unevena}
   }
   \hspace{0.2cm}
\subfigure[Jitter]{
  \includegraphics*[width=7.6cm, height=4.8cm]{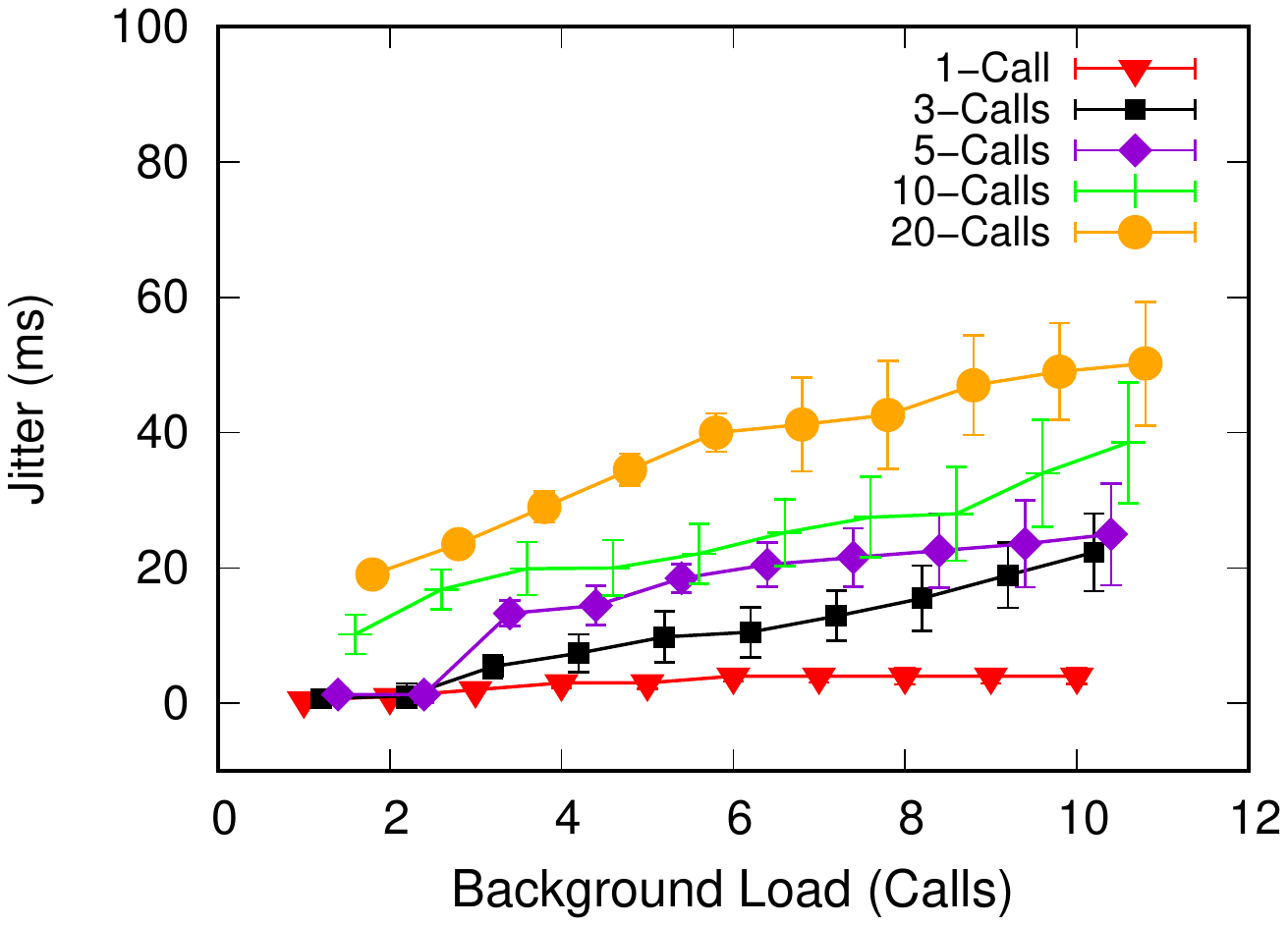}
   \label{Unevenb}
   }
\end{tabular}
\end{center}
    \vspace{-0.1cm}
\captionsetup{singlelinecheck=true} \vspace{-0.25cm} \caption{\label{OutdoorMetrics}
     \footnotesize{Outdoor Testbed Performance Metrics}}
\vspace{-0.5cm}
\end{figure*}

\subsubsection{Lessons Learned}
\label{lessonsLearned__outdoor}
Several visits were performed to the outdoor Fatimah Jinnah Park field site to test various features of the WiMesh system. The initial field trials resulted in several problems that prevented successful execution of the WiMesh system. Results were recorded only for the final field trials. Some of the problems observed during field trials and subsequent improvements to the WiMesh system were:

\begin{itemize}
\item When high antennas were used, they were sometimes unstable due to wind. This problem was mitigated by installing small hard pads on end of pole legs and then embedding these legs in the ground.

\item Ethernet switches sometimes malfunctioned, especially when there was even a slight difference in the applied voltage (applied ${\sim}12$ volts vs. ${\sim}5-9$ volts) normal range. The malfunctioning did not cause the mesh nodes to stop communicating, but rather started a never-ending cycle of disconnections and re-connections which resulted in intermittent connections to server over the mesh. This hardware fault was discovered after much difficulty as the root cause was not obvious;
\item A communication problem was discovered that some mesh nodes were not aware of the existence of the server unless specifically advertised by the server mesh router. This issue was handled by providing Host-Network Association (HNA) announcements in the routing protocol configuration. The scripts were appropriately updated;
\item The range of some of the wireless hardware was extremely poor in outdoor scenarios. More notably, the Ubiquiti PicoStation performed quite poorly in outdoors for distances greater than half a kilometer;
\item Sometimes, a user would be online, however, he would be shown offline. This user offline issue was due to the fact that the periodic update mechanism was flawed whereby each node was required to send data periodically to the server. The socket destruction technique was employed for detecting out of range users to solve this problem;
\item There was previously no mechanism of acknowledging SMS texts, so the sender party had no information whether the text had been delivered or not. This problem was resolved by including notification as soon as message is successfully received at the other end; and
\item When a user sent a video stream request, the sending user did not know whether the destination had accepted the request or not. This was fixed by the sender getting a confirmation from the receiver.
\end{itemize}

\vspace{2mm}
\subsubsection{Testbed Results}
The experiments were repeated under different timings and traffic load conditions. Figure \ref{OutdoorMetrics} shows the performance evaluation graphs for 95\% confidence intervals. It is pertinent to note for all metrics, the results are slightly worse than the indoor case and this is due to several reasons including long outdoor wireless links with fading, interference and even weather conditions. Figure \ref{OutdoorMetrics}(a) shows the Packet Delivery Ratio (PDR) for varying background loads vs. varying number of calls. As the results show, the performance degrades as the load increases, however, the performance still remains acceptable. Figure \ref{OutdoorMetrics}(b) shows the Packet Loss Ratio (PLR) which corresponds indirectly to the PLR and shows similar trend of increasing loss with increased load. Figure \ref{OutdoorMetrics}(c) shows the end-to-end delay which are elevated compared to the indoor scenario. Figure \ref{OutdoorMetrics}(d) shows the jitter observed for the calls, which is again considered acceptable. Overall, the results show that the performance degrades slightly compared to the indoor case, however, given the long wireless links and the outdoor environmental factors, such degradation is expected. 

\subsection{Ziarat Deployment}
WiMesh system was deployed in the Ziarat mountanious region of Pakistan after the Ziarat earthquake disaster \cite{ZiaratEarthquake}\cite{PakistanEarthquake}.  This section shares WiMesh deployment experience, performance evaluation results, and the lessons learned during the Ziarat deployment. 

Ziarat is a remote mountainous region in South-West Pakistan with several villages and small towns scattered over the Suleiman mountain range stretching for hundreds of kilometers. It is sparsely populated and modern amenities such as electricity and telecommunication are sporadically available. Ziarat was hit by a catastrophic earthquake of magnitude of 6.4 in 2008. The WiMesh system was deployed after the earthquake. 

Apart from the problems of disaster and the subsequent rehabilitation, the major problem faced by these villagers was the fact that they had no mechanism of communication with each other and there was no cellular coverage in this region. This was particularly troublesome during an emergency situation where the men working in their gardens and fields could not be reached immediately. Before the WiMesh deployment, the mechanism that they used was that a person had to go physically to the fields and informed them of any emergency situation. However, since the mountain ran across its length and divided the residential areas from the fields, it made accessibility to villagers working in the fields even harder for any emergency situation. Another problem faced by the villages was the lack of communication between the water supply room (generator powered) and the villagers. Since the time for water for the fields was rationed among different villagers, so it would have been immensely helpful for coordination if people could somehow communicate. 

These problems provided an ideal situation and a stringent test for the WiMesh deployment. In particular, two villages in Ziarat, Ahmadun and Gogi, were visited several times for practical testing and final WiMesh deployment.

\subsubsection{Deployment Logistics}
There were several aspects that were unique and challenging in deploying WiMesh to Ziarat. Perhaps the most interesting aspects were rather non-technical. The first mission was to take the villagers in confidence about the strange equipment and techies that would be roaming about. To put things into perspective, a 300 years old village named ``Gogi'' was selected for the deployment, which is entrenched deep inside the mountains in Ziarat with a population of around 2000-3000. It is shouldered on two sides by massive mountains and a semi-paved 20 km road is the only access off the main Ziarat road. The tribal people of this region are known through centuries for their bravery, pride, and hospitality. Fortunately, the lead author had personal contact with some people in the village to gain access for deployment. In addition, an agreement was signed with the Provincial Disaster Management Authority (PDMA), Balochistan Province of Pakistan, to become WiMesh deployment official partner and to help with logistics, regulation compliance, and other aspects of deployment. PDMA's help in overcoming the regulatory compliance was important because the national telecommunication authority has strict regulations over the transmission power/range of outdoor Wi-Fi devices, and the WiMesh deployment plan was to use the maximum power possible in this remote region for optimal communication.

\begin{figure*}
\begin{center}
\label{Ziarat}
\begin{tabular}{c c}

\subfigcapskip 2pt \subfigure[Ziarat Mountain Range]{
  \includegraphics*[width=7.6cm, height=4.8cm]{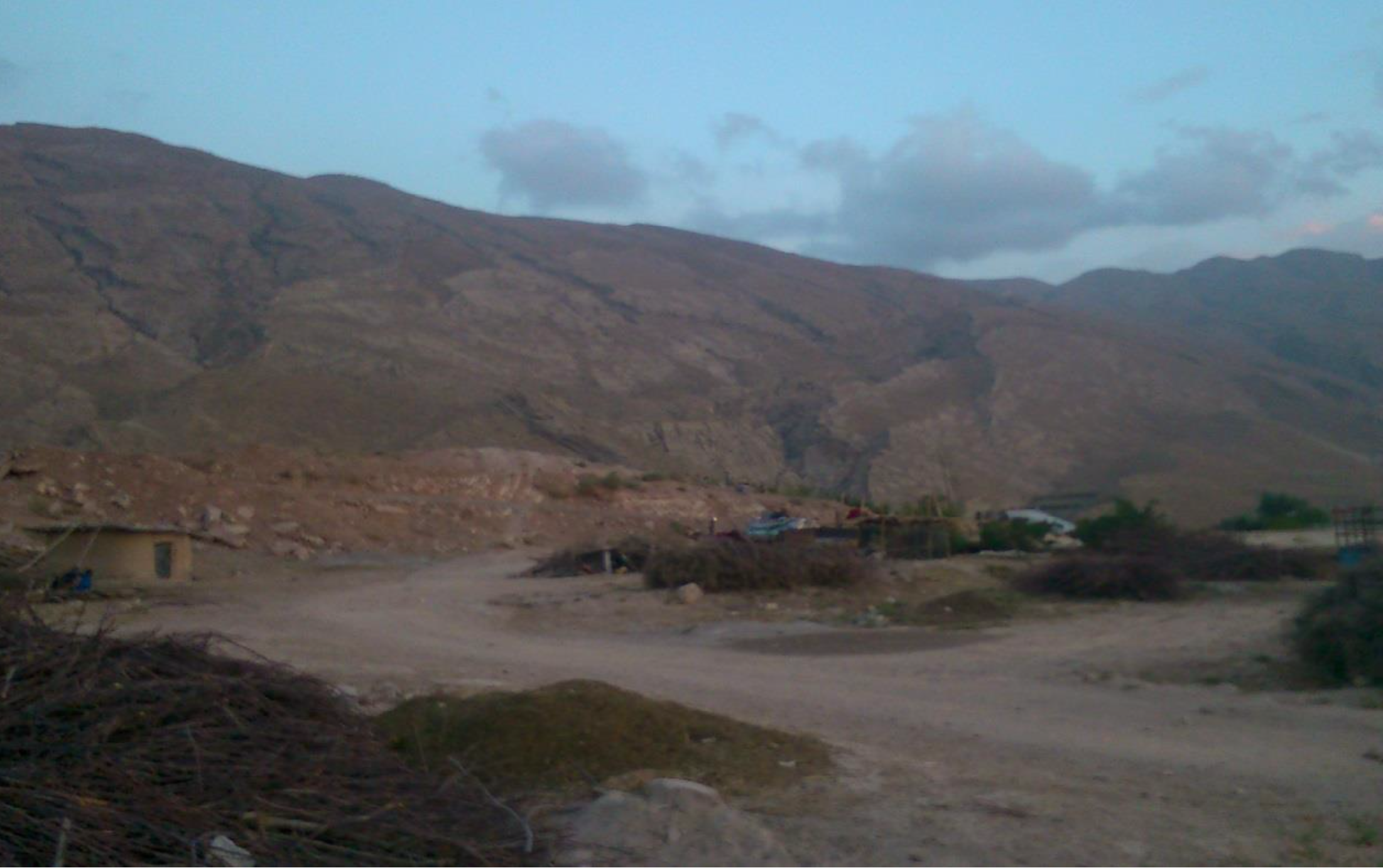}
   \label{ziarat_a}
   }
   \hspace{0.2cm}
\subfigure[WiMesh Deployment Team]{
  \includegraphics*[width=7.6cm, height=6cm]{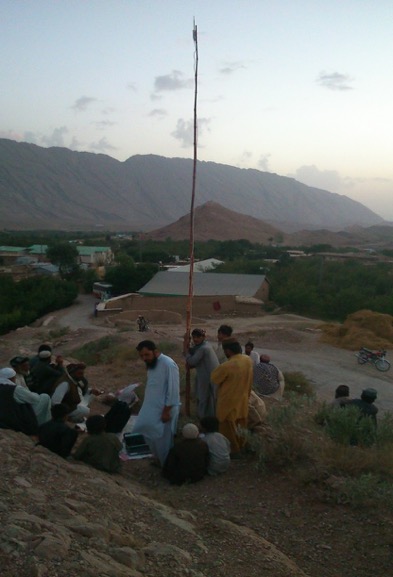}
   \label{ziarat_b}
   }\\
   \subfigcapskip 2pt \subfigure[WiMesh Ready for Deployment]{
  \includegraphics*[width=7.6cm, height=4.8cm, angle=270, origin=c]{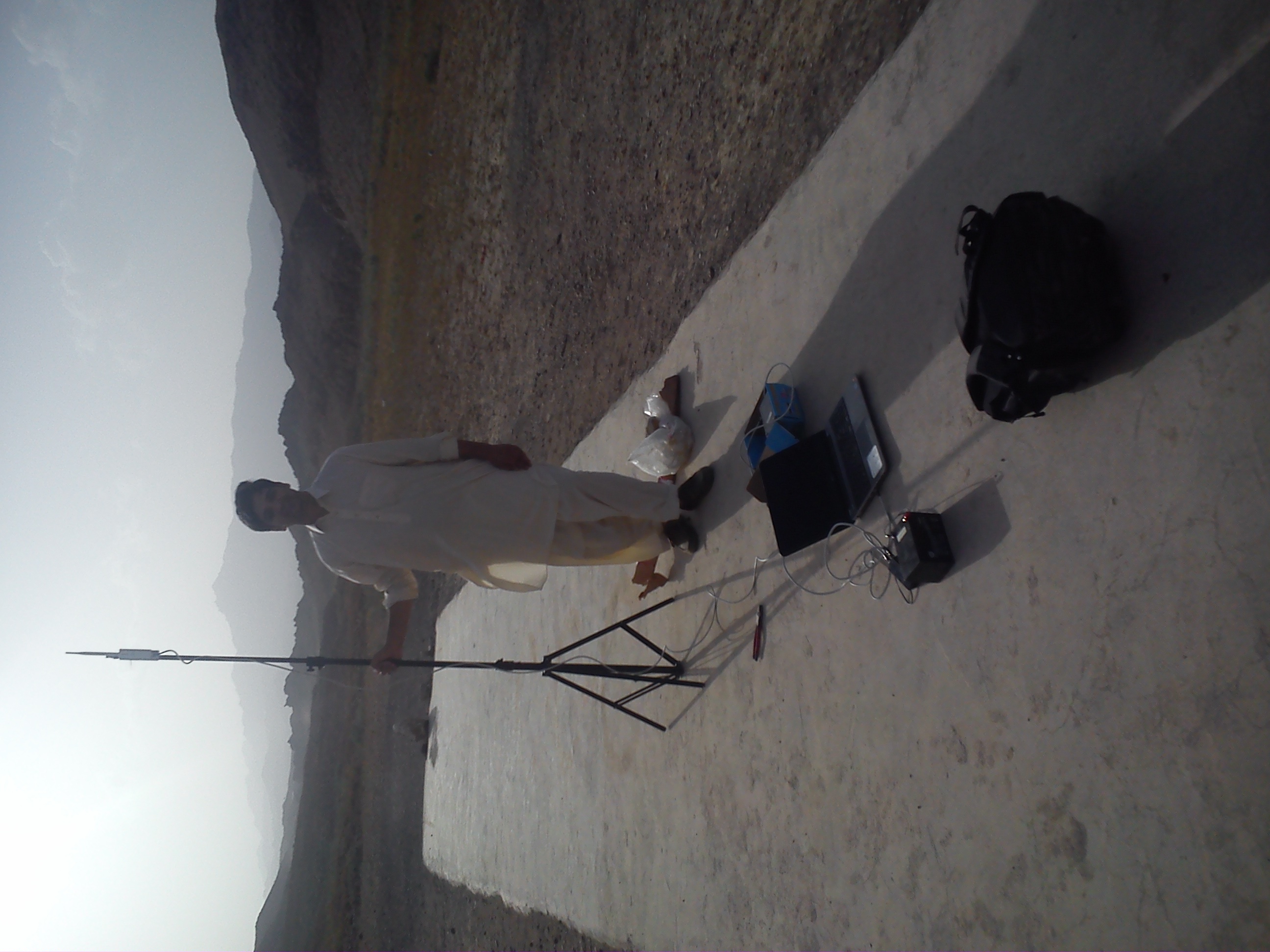}
   \label{ziarat_c}
   }
   \hspace{0.2cm}
\subfigure[Villagers Gathered for WiMesh Demonstration]{
  \includegraphics*[width=7.6cm, height=4.8cm]{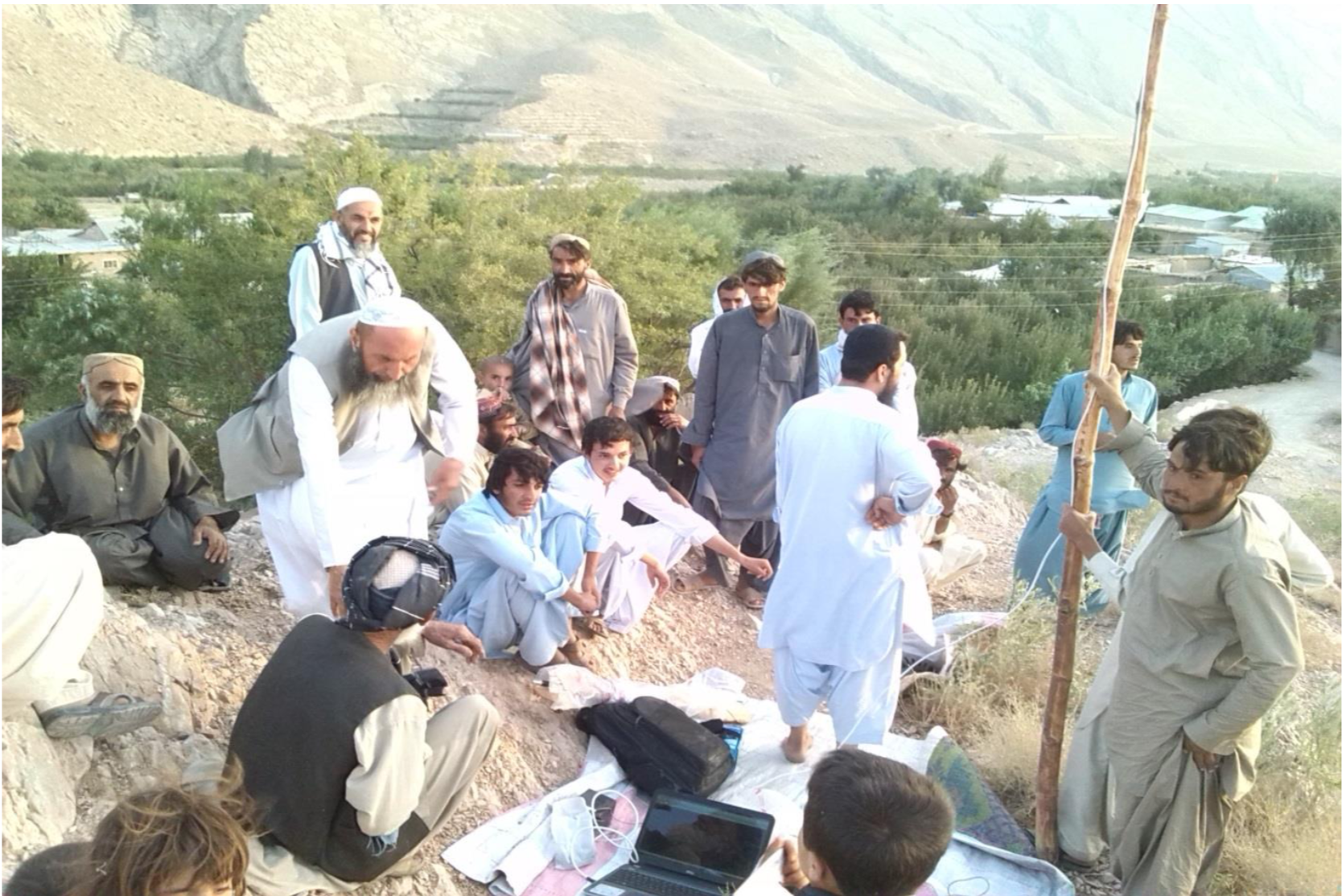}
   \label{ziarat_d}
   }
\end{tabular}
\end{center}
\vspace{-0.1cm}
\captionsetup{singlelinecheck=true} \vspace{-0.25cm} \caption{\label{ziaratdeployment}
     \footnotesize{Ziarat Deployment}}
\vspace{-0.5cm}
\end{figure*}

Information exchange sessions were held where the villagers shared their vows due to the earthquake and, in general, they were provided awareness about the WiMesh system (in layman terms) and how it could help ease communication problems. The village itself has a small mountain dividing the village into two lateral sections with one area covered with hectares of apples, cherry, and apricot farms while the other section contained the housing. Even though, a heavenly town (frequented yearly by the lead author), it is plagued by several problems which have worsened after the earthquake.

The deployment team was informed that one of the problems was that when men worked in the fields, the women back home could not contact them in case of emergencies or just for communication since there is no telecommunication coverage and electricity is rarely available especially after the earthquake. It was demonstrated to the villagers how WiMesh could help by starting with an initial deployment of two mesh nodes nearly 1.5 km apart on top of the small mountains and demonstrated to the utmost joy of the villagers how they could make calls to each other and how solar panels with batteries could be used to power the nodes indefinitely. 

Other logistic concerns included portability of equipment since even the roads inside the village are tough to trek or drive on and it is not easy to carry heavy equipment. 

\subsubsection{Performance Results}
\begin{figure*}
\begin{center}
\label{ZiaratMetrics}
\begin{tabular}{c c}

\subfigcapskip 2pt \subfigure[Ziarat PDR]{
  \includegraphics*[width=7.6cm, height=6.8cm]{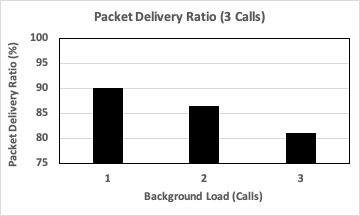}
   \label{ziarat_metrics_a}
   }
   \hspace{0.2cm}
\subfigure[Ziarat Delay]{
  \includegraphics*[width=7.6cm, height=6.8cm]{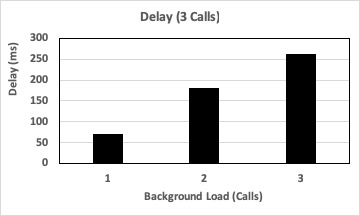}
   \label{ziarat_metrics_b}
   }
\end{tabular}
\end{center}
    \vspace{-0.1cm}
\captionsetup{singlelinecheck=true} \vspace{-0.25cm} \caption{\label{ZiaratMetrics}
     \footnotesize{Ziarat Deployment Performance Metrics}}
\vspace{-0.5cm}
\end{figure*}

Figure \ref{ZiaratMetrics} shows the performance metrics for three calls in the Ziarat deployment. Figure \ref{ziarat_metrics_a} shows the Packet Delivery Ratio (PDR) for varying background loads. As the results show, the performance degrades as the load increases, however, the performance still remains acceptable. Figure \ref{ziarat_metrics_b} shows the end-to-end delay for three calls load.

\subsubsection{Lessons Learned and Challenges}
Some of the main lessons learned during the Ziarat deployment include:

\begin{itemize}

\item The WiMesh team was pleasantly surprised at the extreme hospitality of the villagers and the way they welcomed the team into their homes and their lives. Most of the deployment was straightforward and automated, however, there were still aspects of deployment that needed some manual labour such as hoisting the poles (as shown in Figure \ref{ziarat_b}) and fixing solar panels. Due to the prior agreements, connections, and villagers full support, the villagers were willing to provide any kind of cooperation including providing man-power for the manual labour that was very helpful in the success of the deployment.  

\item The WiMesh team faced several challenges with PDMA partnership mainly due to the fact that it is a pure 
governmental organization with its specific channels and standard operating procedures, some of which can be quite time consuming. Good deployment planning, however, helped overcome some of these limitations. The more serious problem was the fact that during the 19 months of this project, the PDMA Balochistan had seen the change of about five Director Generals. Each time, the Principal Investigator had to work from scratch on developing linkage with the new Director General, explaining to him the motivation and utility of the project and the progress made so far. Some were considerate and helpful, while some were not interested and this required significant effort on part of the Principal Investigator to keep them in loop and motivate them.

\item The plug-and-play nature of WiMesh did make deployment quite easy, as it was realized that any kind of last minute ``tweaks" and adjustments would have been impossible or extremely difficult without this feature. Moreover, sometimes the network needed to be restarted due to energy depletion (explained below) or some technical fault and it was a big advantage that simple villagers could simply press a few buttons and our automated scripts would render the system in a functioning state without any specialized intervention.

\item It was realized that assumption of indefinite power supply through a combination of solar panels and batteries did not hold true since this region had mostly cloudy weather with sparse sunshine throughout the year except for the few months of summer. The region has long and harsh winters and few months of sunny summers. It was learned that solar panels still generate electricity even in the cloudy weather, albeit in significantly smaller quantities.  Therefore, the project team decided to use better quality solar panels and bigger batteries, however, this came at the cost of compromising the cost-effectiveness. In addition, the team modified the routing protocol and metric to further restrict unnecessary routing messages to conserve energy. Despite all these changes, a 24/7 operation could not be guaranteed. Thus, the network was primarily used during the days. However, this limitation had minimal implication as the villagers had a natural lifestyle and slept early around 7pm in winters and 8pm in summers. The plug-and-play nature of WiMesh again helped in making sure that villagers were able to easily restart the network if it went down during the long winter nights.

\item Since, the poles were of varying heights (and sometimes makeshift), the project team resorted to designing the Ethernet cables of varying lengths themselves and installing the RJ-45 connectors. Due to the custom design of the connectors and cables, sometimes these connectors would become loose due to rough handling and caused all kind of connectivity issues. This situation was exacerbated by the harsh deployment scenarios. To add to the complexity, the connectivity issues could be partial or time-varying as well, i.e., instead of a complete disconnection, we could get partial connectivity or high packet losses. Sometimes several hours would  be wasted in diagnosing connectivity issues. This was particularly true for the mesh node hosting the server which could bring down the complete network. An effective and simple solution that eventually solved our problem was to use off-the-shelf Ethernet cables with professionally installed connectors, which do not come loose despite rough handling. This solved most of our connectivity problems. 

\item Due to the success of WiMesh deployment, the villagers requested to further extend the scope of the project and to connect the WiMesh network to a Base Transceiver Station (BTS) several kilometers away. The team investigated and did an initial feasibility analysis for this request. It turned out that there was an open-source software called OpenBTS which offered open-source BTS functionality for remote regions and has been deployed in impoverished areas of Africa \cite{openbts}. The team investigated the complexities of integrating WiMesh with OpenBTS and interfacing with the commercial telecommunications provider. Due to the lack of time and sufficient resources, the team eventually decided to pursue it as a future project.
\end{itemize}

\section{General Insights and Lessons Learnt}

This section reports some of the salient insights, distilled from the project experience in developing and deploying the WiMesh system. A valuable contribution of this work is that different kinds of unexpected causes of failure were documented that may occur when deploying wireless technologies for emergency management, which are expected to benefit researchers undertaking such deployments in the future. In particular, the section describes how these insights add to the body of knowledge and also report where these findings are aligned with or discordant with existing wisdom. This section complements the earlier discussions in Sections \ref{lessonsLearned__indoor} and \ref{lessonsLearned__outdoor} on the salient lessons learned during the experimentation with the indoor and the outdoor testbeds, respectively.

\subsection{Sustainability of Wireless Based Emergency Networking}

To develop a sustainable solution, it is important to factor in the unique requirements of the particular deployment area. It is well known in the ICT for development (ICTD) community that it is a real challenge to keep a rural wireless network running for long-term. Most projects fail to go beyond the pilot stage \cite{surana2008beyond} and techniques such as resource pooling \cite{qadir2016resource} can help. Since WiMesh is low-cost, rugged, and customized for the environment, therefore, the system is still operational in some villages, which supports its sustainability aspect.

\subsection{Power Related Issues}
Apart from the lack of a sustainable financial model, the most common operational cause underlying the lack of sustainability of most wireless networking projects in the rural areas is the high cost of dealing with the poor quality of power, which is a very common issue in rural areas of the developing world \cite{surana2008beyond}. The poor power quality also leads to higher failure rates of components, which means that ``the  real cost of power is not the grid cost, but is the cost of overcoming poor power quality problems'' \cite{surana2008beyond}. This is because low-quality power necessitates an additional burden of deploying additional infrastructure such as the use of power controllers, batteries, and solar-powered backup power solutions.

\subsection{Anticipating Malfunctions and Fault Diagnosis}

For building a sustainable solution, it is important to have the ability to anticipate malfunctions and diagnose faults. A number of authors have written about simplifying and automating fault diagnosis, and interested readers are referred to \cite{surana2007simplifying} and \cite{gabale2013deployments}. Subramanian et al. \cite{subramanian2006rethinking} have shown that low-cost community wireless networks can malfunction in a myriad ways from complete failures (such as hardware board failures, corruption of the flash memory cards, and lightning strikes) as well gradual degradation over time (such as interference from external sources, effects of unreliable power, and rainfall). 

As noted in Sections \ref{lessonsLearned__indoor} and \ref{lessonsLearned__outdoor}, it was also discovered through this project experience the great utility of having a featured monitoring software since it is very common for wireless interfaces on low-cost devices to malfunction (often due to imperfect power supply). This malfunction then leads to knock-on problems such as intermittent performance, cyclic loop of disconnections and reconnections, or routing issues that are hard to troubleshoot back to the offending cause. 

\subsection{Open-Source Community Software Contribution}
We believe that it is necessary for the wireless networking for emergency community that we develop an ecosystem that encourages sharing of insights and the communal reuse of basic building blocks, which practitioners and researchers can use to develop customized solutions more efficiently, more quickly, and with lesser resources. In the same spirit, we have leveraged existing open-source solutions where available, and are releasing the custom-developed source code of the WiMesh system, as well as detailed instructions and usage guidelines. These resources can be accessed publicly at the website of the WiMesh project (\url{https://sites.google.com/site/wikiwimesh/}).

\section{Acknowledgement}
The authors would like to extend their thanks to the National ICT R \& D Fund Pakistan for providing the financial support towards this project.
\section{Conclusion}
This paper presented the WiMesh disaster communication system including its conception, design, implementation and field trials. The WiMesh system leverages the wireless multihop capability of the mesh network and the Wi-Fi accessibility of mobile devices to connect to the nearest mesh node. Due to the nature of their application domain, disaster communication systems need to have specific traits of size, portability, power, cost, customization, ease of installation, wireless coverage, and availability of hardware and software. Some of these traits, such as cost and availability, are especially critical for deployment in developing and third world countries. These features were considered important during the design and development of the WiMesh system.  Furthermore, as discussed in the paper, the WiMesh system meets all of these criteria. The paper provided architectural and design details of the WiMesh system including design trade-offs, pitfalls, and challenges faced during the design, development, and deployment. Extensive indoor and outdoor field trials were carried out to demonstrate the functionality and to collect performance metrics for the WiMesh system. Details of these field trials along with performance metrics results were also presented in the paper. These field trial results demonstrate that the WiMesh system can handle several dozen concurrent VoIP calls without noticeable performance degradation, even in the presence of significant background traffic. There are a number of areas that can be further improved. For instance, 5G technology is making inroads in the wireless ecosystem and it would be important to leverage that strength into the WiMesh solution.

\bibliographystyle{ieeetr}
\bibliography{biblio}
\end{document}